\documentclass[epj]{svjour}

\usepackage{latexsym}
\usepackage{amsmath}
\usepackage{graphicx}
\usepackage[utf8]{inputenc}
\usepackage{url}
\begin{document}
\onecolumn

\newcommand{\ds}{\displaystyle}
\newcommand{\D}{\mathrm{d}}
\newcommand{\I}{\mathrm{i}}
\newcommand{\abs}[1]{\left\lvert #1 \right\rvert}
\newcommand{\EXP}[1]{\mathrm{e}^{#1}}
\providecommand{\vec}[1]{\boldsymbol{#1}}
\newcommand{\abl}[3][\empty]{\frac{\D^{#1} #2}{\D {#3}^{#1}}}
\newcommand{\pabl}[3][\empty]{\frac{\partial^{#1} #2}{\partial {#3}^{#1}}}
\providecommand{\PACS}[1]{\empty}

\title{Reduction of a family of metric gravities} \subtitle{With
  highlights on conservation laws in metric formulations, consistency
  before dynamics, and a fresh view on the unity of Newtonian and
  Einsteinian gravity}
\author{
  Klaus Kassner
 \thanks{Klaus.Kassner@ovgu.de}
}        
%
%

\institute{Institut für Physik,
  Otto-von-Guericke-Universität, PF 4120,
  39016 Magdeburg, Germany}
%
\date{July 25, 2019}
%
\abstract{ A recent proposal by Shuler regarding a postulate-based
  derivation of a family of metrics describing the gravitational field
  outside a static spherically symmetric mass distribution is
  reviewed. All of Shuler's gravities agree with the Schwarzschild
  solution in the weak-field limit, but they differ in the
  strong-field domain, i.e., close enough to a sufficiently compact
  source of the field. It is found that the evoked postulates of i)
  momentum conservation and ii) consistency of field strength
  measurement are satisfied in all metric theories of gravity
  compatible with the Einstein equivalence principle, no matter what
  the form of the metric.  Therefore, they cannot be used, within any
  correct deduction, to \emph{derive} a particular metric. Shuler's
  derivations are based on an inconsistent set of correspondences
  between local and distant quantities. Furthermore, it is shown
  here that out of the family of possible metrics given by Shuler only
  one member, the Schwarzschild metric, satisfies a standard
  relativistic generalization of Newton's law of gravitation,
  suggesting the others to be unphysical.
   } 
\maketitle

\allowdisplaybreaks
\section{Introduction}
\label{sec:intro}
There are a number of reasons why one might wish to derive the metric
around a spherically symmetric mass distribution without using
Einstein's field equations. One is teaching: several appealing
physical results, including those referring to the four classical
tests of general relativity, are derivable from that metric. This may
then be motivating enough for students to tackle the mathematics of
the field equations, whereas plunging into difficult differential
geometry first might rather demotivate them. Another is historical
interest: it is amusing to speculate how Einstein, had he been in the
possession of an exact solution before the field equations, might have
avoided some of the tedious detours (such as the ``Entwurf theory'' of
1913) that he actually took \cite{weinstein18}, before finally
settling on the correct field equations in a race with Hilbert. Third,
a successful shortcut towards the Schwarzschild solution may be
helpful in approaching a deeper understanding of the theory of
gravity, both for the teacher and students.

These were some of the motivations for a series of
papers \cite{kassner15,kassner17a,kassner17b} I wrote in the last few
years, exploring roads to the Schwarzschild metric with as few
postulates as possible beyond the standard ingredients, two of which
come from the ``easy'' sector of general relativity (GR) describing
motion of matter in the gravitational field, viz.~the equivalence
principle and special relativity (SR), the third being the Newtonian
limit (NL), required to apply far enough from matter distributions.
In Ref.~\cite{kassner15}, the single postulate was used that a
static gravitational field should be source-free in vacuum, which
made a certain divergence vanish and provided a relationship between
the two functions that may be chosen freely in a time independent spherically symmetric
metric after coordinates have been fixed.\footnote{The most general
  form of a stationary spherically symmetric metric contains four
  independent functions of the radial coordinate \cite{kassner17a}, of
  which two can be eliminated by requiring, first, the radial
  coordinate of an event to be the length of the circumference,
  divided by $2\pi$, of a circle about the symmetry center through it
  and, second, the time coordinate to be orthogonal to the spatial
  foliation.} This condition was not sufficient to fully determine the
metric,\footnote{Or at least, I was not able to exploit it to that
  effect.} so the known experimental result on the perihelion
precession of Mercury was used to fix the $g_{rr}$ coefficient to
lowest nontrivial order in $1/r$, which then yielded a weak-field
metric that was sufficiently accurate to predict the famous factor of
two between Einstein's 1911 result for light-bending by the Sun and
his final prediction. In the second paper \cite{kassner17a}, another
postulate was introduced, this time a dynamical one, requiring gravity
to not deform the shape of a spherical (scalar) wave about the
symmetry center, in a freely falling frame. This postulate is
based on stipulating mathematical simplicity of a
law of physics. To the large majority of physicists, it may seem less
convincing than the first postulate with its clear physical
contents. While these two postulates were sufficient to derive the
full Schwarzschild metric, the mathematical nature of the second and the
mere fact that \emph{two} postulates were needed rather than a single
one, may have impaired the persuasive power of the approach. This was
remedied in the final article \cite{kassner17b}, in which a single
requirement on the behavior of dust balls in a gravitational field
turned out to enable a derivation of the exact Schwarzschild
metric. Moreover, this requirement is, as I will argue in this
article, nothing but a straightforward generalization, according to
well-accepted rules, of Newton's universal law of gravitation (in
vacuum) to the relativistic case. Conceptually, things cannot get
simpler than this, and the approach is convincing independent of the
question whether to interpret GR formulas as referring
to space-time itself or to tensor fields on some unexplored
background.

Some care was taken in these articles to avoid the fallacies of preceding
attempts at a simple derivation of the Schwarz\-schild metric. 
In particular, the postulates employed apply to the limit of a
parallel gravitational field as well and allow one to thus derive the
Rindler metric, a test that any postulational approach to the
Schwarzschild metric should have to pass \cite{kassner17a,kassner17b}.
Moreover, they are not derivable from the Einstein equivalence
principle (EEP), special relativity and/or the limit of Newtonian
behavior of the gravitational field at infinity. After all, it has
been demonstrated in some mathematical detail that these three
ingredients are insufficient to rigorously justify the Schwarzschild
metric \cite{gruber88}. This should be obvious: the field-theoretic
sector of GR describing the generation of gravity by mass-energy
contains additional physics beyond the matter sector characterizing
motion of particles in an existing gravitational field. The physics of
the latter arises from SR augmented by the EEP, but the solution to the question of
how gravitational fields are generated required additional ideas.

There have been many attempts at straightforward derivations of the
Schwarzschild metric or some of its consequences (such as the correct
description of light bending) in the history of modern physics.  These
were probably motivated by the simple form of the metric in
Schwarzschild coordinates. An early idea by Lenz \cite{sommerfeld52}
and Schiff \cite{schiff60} became quite influential, as modern authors
repeated their mistake \cite{rowlands97,cuzinatto11} in spite of
Rindler's demonstration \cite{rindler68}, via a counterexample, that
Schiff's argument does not work and notwithstanding the mathematical
analysis given by Gruber et al. \cite{gruber88}. In particular, Rindler
shows that while the $g_{tt}$ term of the metric may be inferred from
time dilation (as observable via photon redshift) on the basis of
appropriate clock dropping experiments and that a procedure involving
the dropping of rulers may be made to work for the definition of
lateral coordinates leading to $g_{\vartheta\vartheta}$ and
$g_{\varphi\varphi}$, a similar method for radially oriented rulers
will not produce a result on $g_{rr}$.

The first approach obtaining the Schwarzschild metric via postulates
taking the place of the field equations seems to be due to
Tangherlini \cite{tangherlini62}. While it was favorably
received by Sacks and Ball \cite{sacks68} in yet another rebuttal of
Schiff-like arguments, Rindler later demonstrated \cite{rindler69} that
one of postulates of Ref.~\cite{tangherlini62} (termed ``strong
version of the principle of equivalence'' by its author) was in fact a
coordinate postulate that failed when applied in attempting to derive
the Rindler metric from it. Hence the postulate was not convincing, as
its validity seemed to be restricted to the spherically symmetric
case.

A remarkably concise postulate-based derivation of the Schwarzschild
metric has been given recently by Dadhich
\cite{dadhich15}. Unfortunately, one of his postulates is without
plausible foundation in the absence of the field equations. To wit, he
requires the ``acceleration of a photon'' $\ddot r$ to be zero on
radial geodesics. Here, $r$ is the circumferential radial coordinate
and the double derivative is with respect to an affine parameter on
null geodesics. If, instead, the metric had been written in isotropic
form (see, e.g., \cite{mueller10}) with a different radial coordinate
(often denoted as $\rho$), the radial acceleration of a photon would
have been nonzero. So the postulate remains unconvincing as long as no
argument is given, why it is the circumferential coordinate that
should lead to vanishing coordinate acceleration. Dadhich uses this
postulate to derive $g_{tt} g_{rr}=-1$ in the Schwarzschild metric. In
fact, it has been shown before that, in a static metric, a sufficient
and necessary condition for this property to hold is that the radial
coordinate is an affine parameter on a radial null geodesic
\cite{jacobson07}. 
Hence, it holds true in the Schwarzschild metric but its
analog does not hold in the standard form of the Rindler metric.
Dhadich's approach thus fails the test of Rindler's
counterexample as well.

Finally, Shuler published, in this journal \cite{shuler18}, an alleged
derivation of a ``family of metric gravities'', including the Schwarzschild
solution, from conservation principles instead of the field
equations. To obtain the Schwarzschild metric proper, he requires in
fact three postulates, which looks like a step backward in comparison
with Ref.~\cite{kassner17b}, where just a single one is needed.

The purpose of the present paper is to show, on the one hand, that out
of this family only a single member satisfies a straightforward
generalization of Newton's law of gravitation to
the relativistic case, using well-known rules of how to transform
kinematic laws from their prerelativistic form into a correct
relativistic version. No surprise, the surviving member of the family
is just the one predicted by GR. On the other hand, I would like to
point out that the two postulates on which Shuler bases the central
part of his derivation, i.e., conservation of momentum and consistency
of field strength measurement are satisfied in \emph{any} metric
theory due to the equivalence principle and thus do \emph{not}
constrain the metric. They are not independent postulates but
incorporated in the Lagrangian equations of motion following from the
metric.

That Shuler nonetheless obtains a condition for the coefficients of
the metric, is due to his use of inconsistent expressions connecting
distant with local quantities. This leads him to the erroneous belief
that the inverse proportionality of $g_{tt}$ and $g_{rr}$ is a
physical principle rather than a result of coordinate choice, in spite
of the mathematical analysis of Ref.~\cite{jacobson07} proving
the contrary. 

A postulate with physical content used by Shuler (beyond the EEP, SR
and NL) is the requirement of a form of Gauss's law (effectively
stating that the field has no sources in vacuum). This indeed
constrains the metric by establishing a relationship between the two
radial functions appearing in it.  However, Shuler discards
this -- most sensible -- postulate in favor of far less convincing
alternatives to obtain a family of metrics.

The rest of this article is organized as follows. In
sect.~\ref{sec:cons_laws}, the meaning of conservation of energy and
momentum in the framework of a metric description is discussed. It is
shown that these conservation laws lead to the Lagrangian equations of
motion, aka geodesic equations, but do not constrain the metric in any
way. Section \ref{sec:CFS} deals with the question on how certain
quantities pertaining to local phenomena at one place in the metric
can be defined operationally for a distant observer (at some other
place in the metric). In the absence of uniqueness, there is some
choice in doing so, but a sanity requirement is that any set of
definitions must be internally consistent. It is argued that Shuler's
set is inconsistent. Moreover, consistency of field strength
measurement is shown not to lead to constraints on the
metric. Shuler's idea that $g_{tt} g_{rr}=-1$ reflects a property of
space-time rather than one of coordinates is refuted. Finally, it is
demonstrated in sect.~\ref{sec:gen_newton_grav} that a particular form
of Newton's law of gravitation may be generalized to a
relativistically valid form in a standard straightforward manner. This
generalization yields a criterion to pick out a single member of
Shuler's family of gravities. Section \ref{sec:conclusions} summarises
and concludes the paper.

\section{Conservation laws}
\label{sec:cons_laws}

Shuler \cite{shuler18} mentions Noether's theorem implying energy
conservation as a consequence of the homogeneity of time and momentum
conservation as a consequence of the homogeneity of space. He then
talks about choosing homogeneous observer coordinates, albeit without
considering whether those are possible at all. Clearly, there are
situations where no homogeneous (global) coordinates can be
introduced. The surface of a sphere is a homogeneous (and isotropic)
two-dimensional space. Nevertheless, no finite coordinate patch on a
sphere is homogeneous. The line element on a sphere with radius $a$,
written in standard spherical coordinates,
\begin{align}
  \D \ell^2= a^2\left(\D\vartheta^2+\sin^2\vartheta \D\varphi^2\right)\>,
\end{align}
is inhomogenous in the coordinate $\vartheta$ and still describes a
homogeneous space. So whether space is homogeneous or not is not a
question of the homogeneity of coordinates.  Homogeneity means invariance
under spatial translations and is a coordinate independent property.

It is then a somewhat bizarre claim by Shuler that if the
circumferential radial coordinate $r=C/2\pi$ is used as one of the
coordinates in a spherically symmetric metric, the latter will be
homogeneous in $r$. This is not even true for flat space! Let us write
out the Minkowski metric in Cartesian and in spherical coordinates:
\begin{align}
  \D s^2 &= -c^2 \D t^2 +\D x^2 +\D y^2 +\D z^ 2 =  -c^2 \D t^2 + \D r^2 + r^2\left(\D\vartheta^2
           +\sin^2\vartheta \D\varphi^2\right)\>.
           \label{eq:minkowski_metric}
\end{align}
A Lagrangian describing the motion of a free particle in this metric
can be obtained setting $L=\left({\D s}/{\D\tau}\right)^2$, where
$\tau$ is the proper time of the particle. 
Multiplication of a Lagrangian by a constant produces a new valid
Lagrangian and $L'=\frac{1}{2}m\left({\D s}/{\D\tau}\right)^2$ is a
form that reduces, in the nonrelativistic limit, to the classical
Lagrangian up to an additive constant (see
Ref.~\cite{kassner15}).  We read off \eqref{eq:minkowski_metric}
that the Lagrangian does not depend on either $x$, $y$ or $z$, hence
the canonical momenta $p_x$, $p_y$, and $p_z$ are conserved and since
these are the Cartesian components of the momentum $\vec p$, a free
particle will retain its initial momentum in this metric. Momentum is
conserved.

However, we also read off that the Lagrangian \emph{does} depend on
$r$ explicitly, meaning that the radial momentum component
$p_r=\vec p\cdot\vec{e}_r$ is \emph{not} a conserved quantity, in
general.
($\vec{e}_r=\sin\vartheta\cos\varphi
\,\vec{e}_x+\sin\vartheta\sin\varphi \,\vec{e}_y+\cos\vartheta \,\vec{e}_z$
is a unit vector in the radial direction.) Indeed, this can be easily
verified by way of an example. Consider a particle moving parallel to
the $z$ axis (but not along it). Its momentum may then be written
$\vec p=p \,\vec{e}_z$. Since momentum is conserved, $p$ is constant
in time. We have $p_r = p \,\vec{e}_z\cdot\vec{e}_r = p \cos\vartheta$
and the angle $\vartheta$ varies, as the particle moves from
the region $z<0$ to $z>0$, from a value larger than $\pi/2$ to a value
smaller than $\pi/2$. Hence $p_r$ is not conserved, and this is so,
precisely, because space is \emph{not homogeneous} in $r$ (i.e., the
metric is not independent of $r$).

If we treat gravity in a Newtonian framework as a one-body problem,
i.e., we consider how a test body moves in the gravitational field of
a much more massive one, space is clearly not homogeneous, as we have
a gravitational field, and fields are properties of space, by
definition. Of course, the momentum of such a test body will change
under the gravity of the massive body. While we may still state that
the total momentum of the test body and the gravitating one will be
conserved (if the ensemble is embedded in homogeneous space) and a similar
statement would be true in the general relativistic framework, this is
not of relevance for the question of momentum conservation of the test
body alone. If Shuler nonetheless speaks of momentum conservation, he
must have something different in mind. As it happens, people often
refer to the momentum balance equation $\vec F={\D \vec p}/{\D t}$
(i.e., Newton's second axiom) as momentum conservation, and this might
also be Shuler's meaning. A relativistic statement of this would seem to 
require us to specify the relativistic meaning of force.

Actually, here we are in a fortunate situation due to the fact that
relativists prefer an interpretation, in which gravity is \emph{not} a
force. If there is no force, we must have some sort of momentum
conservation. And indeed, the equivalence principle reveals this (plus
energy conservation) to be true and, consequently, provides us with
the full set of equations of motion in a given metric. The EEP says
that a sufficiently small freely falling system is a local inertial
system, in which the physics of SR applies. That is, the motion of
test particles not subjected to additional forces is governed by the
metric \eqref{eq:minkowski_metric} and hence, momentum conservation
applies. However, the momentum will change as soon as the particle has
moved far enough for the local inertial system to lose its validity or
after enough time has passed for that to happen (as locality is to be
required not only in space but also in time). This momentum change is
due to the fact that in the new local inertial system that may be
constructed around the particle at a later time, momentum will have a
different value, for momentum typically changes under a change of the
frame of reference.\footnote{The new frame usually has a different
  velocity from the old one.}  Momentum is a conserved quantity but
not an invariant.

Now we can see immediately, how the equations of motion in a metric
follow from the equivalence principle and local energy and momentum
conservation.  Local energy and momentum conservation imply that the
Lagrangian from the Minkowski metric \eqref{eq:minkowski_metric} (or
one equivalent to it) governs the motion. The existence of a local
frame of reference in which this applies is secured by the EEP. Since
Lagrangian equations of motion are derivable from Hamilton's
principle, which is coordinate free, they take the same general form
in whatever coordinates we may choose. So if we transform back to the
global coordinates describing the metric in an extended patch of
space-time, the equations of motion must still be the Lagrangian
equations of motion (following from the form of the metric in the
global coordinates). This transformation is trivial, as we constructed
our Lagrangian as an invariant (both $\D s$ and $\D \tau$ are
invariants) under coordinate transformations, so it is still given by
$\left({\D s}/{\D\tau}\right)^2$, now with the line element written in
terms of the global metric. The equations of motion so obtained are
the geodesic equations \cite{gron07}.

Hence, the geodesic equations are a direct consequence of energy and
momentum conservation in local freely falling systems, due to the
equivalence principle, for \emph{any form of the metric}. Therefore,
local energy or momentum conservation cannot constrain the form of the
metric.

It might be added that the distinction between momentum and proper
momentum made by Shuler is not particularly useful. Proper momentum
would have to be defined either as the spatial part of four-momentum,
in which case it would just be momentum, or as the momentum of an
object in its own, co-moving, frame of reference, in which case it
would always be zero.

\section{Consistency of field strength measurement}
\label{sec:CFS}

Shuler's second postulate is consistency of field strength
measurement, abbreviated CFS. In words, it requires, reasonably, that
a force, applied to a tether to counter any  weight
at the bottom of the tether must be equal to the force that we obtain
when using the tether as a measuring device, exploiting known
relationships between the energy at the bottom end of a tether and its
perception at the top. The tether argument was introduced in
Ref.~\cite{kassner15} to avoid (or solve) the problem of comparing
radial lengths at different radii in a spherically symmetric
metric.\footnote{Whereas I invented this argument for myself, it is so
  simple that there can be little doubt that similar arguments were
  developed many times before.}  So the requirement is entirely
rational, but the question arises immediately whether it should not be
automatically satisfied in any sensible theory. It should therefore
not be necessary to ask for it in a separate postulate. Stated
differently, a theory that is inconsistent at the kinematic level and
needs a dynamic law (in the form of an additional postulate) to repair
this, is not well-formulated to begin with.

\subsection{How to measure distant properties?}

To discuss Shuler's approach, we need some notation. Let us write the
line element for a static spherically symmetric space-time as
\begin{align}
  \D s^2 &= g_{tt}(r)\, c^ 2 \D t^2 + g_{rr}(r)\, \D r^2 + g_{\vartheta\vartheta}\, \D\vartheta^2
           + g_{\varphi\varphi}\,\D\varphi^2= g_{\mu\nu}\, \D x^\mu\D x^\nu \quad (\mu,\nu=t,\,r,\,\vartheta\,,\varphi)\>,
           \label{eq:spherically_symm_metric}
\end{align}
where $g_{\vartheta\vartheta}=r^2$,
$g_{\varphi\varphi}=r^2\sin^2\vartheta$, off-diagonal elements are
zero, and we use the Einstein summation convention in the second
formula. Note that $\D x^t=c\, \D t$, so our $g_{tt}$ differs by a
factor of $c^{-2}$ from Shuler's, who introduces additional notation,
setting $\varsigma = 1/\sqrt{-g_{tt}}$ and $\rho_{rr}=\sqrt{g_{rr}}$.

Then Shuler writes [his equations (3)] a number of ``coordinate
transformations'' for locally measured quantities at some point $r$
($\vartheta$, $\varphi$) in the metric, allegedly relating them to the
``observer reference'' far from the point $r$.

Clearly, this is a misnomer. A (passive) coordinate transformation is
just a relabeling of the coordinates associated with events in
space-time. It does not affect the object to which coordinate labels
are attached. Also, it is \emph{local}, i.e., the transformation
describes how the coordinates of a particular point in space-time are
changed going from one labeling to the other. So it cannot be used to
connect \emph{distant} points in space-time.

Active coordinate transformations describe some motion of objects,
e.g., a rotation. In Euclidean space, we are entitled to believe that
the object does not change its properties under this motion, due to
the symmetries of the space. Objects are freely movable. This is still
true in spaces of constant curvature. As soon as the curvature is
different at different points, the motion of an object may affect its
geometric properties.  Stresses may build up, to which the object has
to react and, therefore, we cannot be sure it will remain the same
after the transformation. Doing experiments (e.g., rotating objects
consisting of different materials in the same way), we may find out
about such effects and quantify them. Shuler's relationships, to which
no operational meaning is given, are not of this type either. So they are
\emph{not} coordinate transformations.

A more precise idea about the meaning of these relations is gleaned by
inspecting them more closely. Most often, their right-hand side contains
a local quantity such as a time or length interval, a velocity or an
acceleration at some position in the field, to which for definiteness I
will assign the radial coordinate $r_1$. For these quantities, there
exist well-defined measuring procedures using devices that can be set
up in close vicinity of the object to be measured. The measurement
result will, within the accuracy of the device, not depend on the kind
of apparatus used.\footnote{The velocity of a billiard ball rolling
  across a table could be measured using a high-speed camera to take
  pictures at well-defined time intervals or by two light barriers
  connected to a stop watch or else by reflecting sound from the ball
  and measuring the frequency Doppler shift. All well-designed methods
  should give the same velocity.} Moreover, we have clear recipes of
expressing the results in terms of the coordinates appearing in
\eqref{eq:spherically_symm_metric}. For example, a radial velocity is
given by the increment of radial proper length element divided by the
proper time of the coordinate stationary observer at $r_1$:
\begin{align}
  v_{\text{rad}} = \abl{\ell_{\text{rad}}}{\tau} = \frac{\sqrt{g_{rr}(r_1)}\D r}{\sqrt{-g_{tt}(r_1)}\D t}\Bigr\vert_{r_1}
  =\sqrt{\frac{g_{rr}}{-g_{tt}}}\, \abl{r}{t}\, \Bigr\vert_{r_1}\>.
  \label{eq:def_vrad}
\end{align}
On the left-hand side of Shuler's equations, we have what these
measurements should correspond to for the
distant observer (whom I will place at $r_2$). This observer does not
have any local device near the object to be measured nor would it be
of any use to him without a means of reading off the measurement
result from the distance.  A crucial point is now that there is no
\emph{unique} way of relating or connecting all ``measurements at a
distance''\footnote{Scalar quantities are unproblematic, but most
  measurements refer to more complicated entities. Energy, for example
  is not a scalar but a component of a four-vector.} made by the
observer at $r_2$ with the local measurements made by the observer at
$r_1$.\footnote{Certain observations such as looking through a
  telescope or taking the spectrum of light from the distant object
  may obviously be made from a distance. But their translation to
  quantities such as velocities or accelerations is not immediate.}
According to Bunn and Hogg \cite{bunn09} ``\emph{the inability to
  compare vectors at different points is the definition of a curved
  space-time}'' and this applies to velocity (four-)vectors as well.

To clarify this issue of non-uniqueness a little, let me give a
well-known example from cosmology.  Consider the recession velocity of
sufficiently distant galaxies. Whether this velocity is superluminal
or not, is a matter of interpretation.  If we define that velocity as
the change of the proper distance of the galaxy\footnote{As obtained by adding up
  the proper distance increments of local coordinate stationary
  intermediate observers.}  from the Milky Way per cosmological time,
then superluminal velocities will result and we will find that all
galaxies with a redshift exceeding 1.46 move away from us faster than
light, which however does not impede us from seeing
them \cite{davis04}. One advantage of this definition is that it allows
us to calculate (or define!) the size of the current universe. On the
other hand, if we interpret the recession velocity in terms of
parallel transport of the velocity four-vector of the galaxy along the
null geodesic taken by its light to reach us, i.e., if we parallel
transport this vector to the position of the Milky Way and then
extract the three-velocity from it, we will always find a subluminal
velocity. This has the advantage that redshift and recession velocity
are connected via the standard special relativistic Doppler
formula \cite{bunn09}, so this velocity is easily measurable by
evaluation of the redshift.

Equipped with this understanding, we must interpret Shuler's set of
relationships (3) mostly as \emph{definitions} of their left-hand
sides giving properties ``as seen by the distant observer''.  Shuler
does not give any operational justification of these definitions,
hence some checking is necessary to see whether they are meaningful.

His first equation, not yet a comparison between distant and local
quantities, is
\begin{align} \D t = \varsigma \D\tau\>,
  \label{eq:global_t_of_proper_t}
\end{align}
i.e., $\D\tau = \sqrt{-g_{tt}}\D t$. This merely relates a proper time
interval of a coordinate stationary observer to an interval of the
global coordinate time, both pertaining to the same pair of events. Hence,
it is an equality between coordinate differentials evaluated at
the same location. Nevertheless, it is possible to use this to
establish a connection between local and distant times on the basis of
``slices of simultaneity'' defined by the metric
\eqref{eq:spherically_symm_metric}.  This works as follows. To enable
a comparison of time intervals at positions $r_1$ and $r_2$, intersect
the world lines of the two observers at these positions by (3D)
spatial sections corresponding to global times $t_A$ and
$t_B$.\footnote{This is always possible outside the event
  horizon. What is important is that the two world lines are
  intersected by the same surfaces of constant time, so only two of
  these are necessary to delimit intervals on both lines. } The
intervals of global time corresponding to the so-defined pieces of the
world lines are $\Delta t_1= t_B-t_A = \Delta t_2$. Making that
infinitesimal, we obtain
\begin{align}
  \frac1{\sqrt{-g_{tt}(r_1)}}\,\D \tau_1 = \D t_1 = \D t_2 =  \frac1{\sqrt{-g_{tt}(r_2)}}\,\D \tau_2 
  \qquad  \Rightarrow 
                      \qquad \D \tau_2 &= \frac{\sqrt{-g_{tt}(r_2)}}{\sqrt{-g_{tt}(r_1)}} \D \tau_1\quad  \label{eq:comp_prop_times}
   \\                             \underset{r_2\to\infty}{\to} \quad\>\>  \D t_2
                                       &= \frac1{\sqrt{-g_{tt}(r_1)}} \D \tau_1=\varsigma_1\D\tau_1\>.
   \nonumber                                     
\end{align}
Dropping the labels 1 and 2, the relationship becomes identical in
form to \eqref{eq:global_t_of_proper_t}. But it \emph{means} something
\emph{different}. Confusion about the meaning of coordinate
differentials is a major source of errors in attempts by
non-professionals to prove or disprove  a thing about relativity. Equation
\eqref{eq:global_t_of_proper_t} compares two times at the \emph{same}
position $r$, whereas in eq.~\eqref{eq:comp_prop_times} times at
different positions are related to each other. It would therefore
actually be better \emph{not} to drop the position labels 1 and 2.

In principle, the relationship \eqref{eq:comp_prop_times} is not
unique -- different global time coordinates produce different
simultaneity relationships, and not all of them will lead to the same
value of $g_{tt}(r_1)$ in \eqref{eq:comp_prop_times}. However, a few
requirements such as stationarity of the metric plus a choice of
spatial coordinates that make coordinate stationary observers the same
as those in the metric \eqref{eq:spherically_symm_metric} will suffice
to fix the ratio.  It is the same, for example, in the Schwarzschild
\cite{schwarzschild16a} and Painlevé-Gullstrand
\cite{painleve21,gullstrand22} forms of the metric.  Moreover, there
is a physical experiment establishing this ratio of local and distant
times via the redshift of photons.\footnote{Corresponding time
  intervals at the emitter and absorber of an electromagnetic wave
  train are, of course, not determined by coordinate
  simultaneity. Rather their beginnings and ends are determined by the
  null geodesics connecting the pairs of initial and final emission
  and absorption events of the sequence of photons
  \protect\cite{harvey06}. The time dilation factor resulting from
  this is, however, the same in a stationary metric as the one
  obtained from the simultaneity relation. } So this relationship is
definitely acceptable.

While simultaneity slices are based on a coordinization, the resulting
comparison between times at a distance arguably is not purely
abstract.  In fact, Rindler gives, in his book on GR \cite{rindler01},
a procedure for the operational (though not practical) realization of
the coordinate grid corresponding to a stationary metric, using rigid
rods (outside of matter) and clocks that are running fast (or slow)
with respect to standard clocks, by a factor that may depend on
position but is constant in time.

Next, Shuler has
\begin{align}
  \D \ell = \rho_{rr} \D r = \sqrt{g_{rr}} \,\D r\>.
  \label{eq:proper_length_of_r}
\end{align}
Again, this is, in the first place, a relationship between quantities
near the same space-time position -- a proper distance interval
$\D \ell$ and a coordinate interval $\D r$, both at radius $r$. To
achieve a similar relationship between distant observers as for times,
we might try to implement
\begin{align}
  \frac1{\sqrt{g_{rr}(r_1)}}\,\D \ell_1 = \D r_1 = \D r_2 =  \frac1{\sqrt{g_{rr}(r_2)}}\,\D \ell_2 \qquad
  \Rightarrow \qquad \D \ell_2 &= \frac{\sqrt{g_{rr}(r_2)}}{\sqrt{g_{rr}(r_1)}} \,\D \ell_1\quad  \label{eq:comp_prop_lengths_equal_dr}
   \\                             \underset{r_2\to\infty}{\to}\quad\>\>  \D r_2
                                       &= \frac1{\sqrt{g_{rr}(r_1)}} \,\D \ell_1=\D \ell_1/\rho_{rr}\>,
                \nonumber                         
\end{align}
i.e., just as we equated the global $t$ coordinates of world lines of
distant observers, we would now have to equate $r$ values of
corresponding ends of a spatial (i.e., spacelike) line segment through
each observer. However, \emph{this does not work}, as the
(hyper)surfaces of constant $r\,$ in the neighborhood of observer 1 do
not extend to observer 2 and vice
versa. 
Rather, one with a smaller radius is
completely contained in another with a larger radius.  Hence, while $r$ is a global
coordinate, the $r$ values of the distant observer have no definite
relationship to those of the local one. To delimit $\D r_1$ and
$\D r_2$, we need four surfaces, we cannot make do with two. This is a
relevant difference between the time and the radial coordinates.

On the other hand, Shuler does not use
\eqref{eq:comp_prop_lengths_equal_dr} anyway. His connection between a
distant and a local radial velocity,
\begin{align}
  v_{radial} = v_{r\text{-}radial} \,\rho_{rr} /\varsigma
\end{align}
would read, in our notation, 
\begin{align}
  v_{\text{rad}}(r_2) &= \abl{\ell_2}{\tau_2} \underset{r_2\to\infty}{=}   \abl{r_2}{t_2} 
                        = v_{\text{rad}}(r_1) \sqrt{g_{rr}(r_1)}  \sqrt{-g_{tt}(r_1)} = \abl{\ell_1}{\tau_1} \sqrt{g_{rr}(r_1)}  \sqrt{-g_{tt}(r_1)}
                        \underset{\eqref{eq:def_vrad}}{=} g_{rr}(r_1) \abl{r_1}{t_1}\>,
\end{align}
The radial
prefactor gets squared when we express the measured velocity by the
coordinate velocity (whereas the temporal one is canceled out). This
certainly does not look right. There seems to be one prefactor
$\sqrt{g_{rr}(r_1)}$ too many and this may be due to a confusion
between coordinate velocities and measured velocities. Clearly, the
observer at $r_1$ cannot measure $\D{r_1}/\D{t_1}$ (nor
$\D{r_1}/\D{\tau_1}$) by only local means, because to establish the
difference $\D r_1$, she has to measure two circumferences, which is a
nonlocal operation, whereas $\D\ell_1$ is what is directly measured by
local rulers.

Hence, Shuler's relationships between distant and local measurements
do not seem convincing. At least this author cannot find a rational
way of justifying them.\footnote{Effectively, Shuler employs
  $\D r_2 = \rho_{rr} \,\D\ell_1$, which looks incompatible
  with \eqref{eq:proper_length_of_r}.}  Nevertheless, since they may
be viewed as \emph{definitions} of the distant measurement results,
pointing out the absence of an obvious operational realization is not
sufficient. There might still be a counterintuitive one, easily to be
overlooked. What may be done, however, is to rule them out on the
grounds that they form an inconsistent set. This will be shown below.

But then it will be necessary to develop our own \emph{consistent} set of
rules establishing measurements at a distance. On the way, it will
turn out useful to calculate the proper acceleration of a coordinate
stationary observer in the metric
\eqref{eq:spherically_symm_metric}. The tether argument will be
outlined and exhibited to give the same proper acceleration and a
useful definition of force at a distance. This will then be used to
demonstrate that CFS is automatically satisfied in any
metric.\footnote{The demonstration will be restricted to the
  spherically symmetric metric, but the result should be
  generalizable.}

\subsection{Accelerations and forces in the spherically symmetric metric}
The four-acceleration $\vec{a}$ is given by
\begin{align}
  a^\lambda = \abl[2]{x^\lambda}{\tau} + \Gamma^\lambda_{\mu\nu} \abl{x^\mu}{\tau} \abl{x^\nu}{\tau}\>.
\end{align}
For a freely falling particle, all its components are zero and these
equations reduce to the geodesic equations for the coordinate
accelerations. With a coordinate stationary particle (or observer) instead, $x^\lambda$ is
constant for $\lambda =r,\,\vartheta,\,\varphi$, so the above equations simplify to
\begin{align}
\begin{aligned}
  a^t &= c\abl[2]{t}{\tau}+ \Gamma^t_{tt}  c^2\left(\abl{t}{\tau}\right)^2\>, \\
  a^\lambda&=  \Gamma^\lambda_{tt}  c^2 \left(\abl{t}{\tau}\right)^2,\qquad \lambda= r,\vartheta,\varphi\>.
\end{aligned}
\end{align}
There is no need to evaluate $a^t$ from its formula, because we know
that the four-acceleration must be perpendicular to the four-velocity
$u^\lambda$, which for a coordinate stationary observer has only a
non-vanishing component along the time direction, i.e., $u^\lambda=0$
for $\lambda=r,\,\vartheta,\,\varphi$.\footnote{$u^t$ must be nonzero
  then, since the magnitude of the four-velocity is a non-vanishing
  constant ($c$).}  Hence, $a^t$ must be zero. The fastest way
to obtain the needed Christoffel symbols $\Gamma^\lambda_{tt}$ is to
read them off the Lagrangian equations of motion
\begin{align}
  \abl{}{\tau} \frac{\partial L}{\partial \dot{x}^\mu}-  \pabl{L}{{x^\mu}}= 0\>,
\end{align}
where an overdot means differentiation with respect to proper time and
the Lagrangian is $L=\dot{s}^2$, as before, i.e.,
\begin{align}
  L = g_{tt} c^2 \dot{t}^2 + g_{rr}\dot{r}^2+r^2 \left(\dot\vartheta^2+\sin^2\vartheta\dot\varphi^2\right)=-c^2\>.
  \label{eq:lagr_mparticle_spher_symm}
\end{align}
For $\mu=r$, we have
${\partial L}/{\partial \dot{r}} = 2 g_{rr} \dot{r}$ and obtain, after
division by $2 g_{rr}$ (a prime denotes a derivative w.r.t. $r$)
\begin{align}
  \ddot{r} -\frac{g'_{tt}(r)}{2g_{rr}(r)} c^2\dot{t}^2 &+   \frac{g'_{rr}(r)}{2g_{rr}(r)} \dot{r}^2-\frac{r}{g_{rr}(r)}\left(\dot{\vartheta}^2+\sin^2\vartheta \dot{\varphi}^2\right)=0\>,
\end{align}
from which we can read off $\Gamma^r_{tt}$ as well as $\Gamma^r_{rr}$,
$\Gamma^r_{\vartheta\vartheta}$, and $\Gamma^r_{\varphi\varphi}$ (and
see that $\Gamma^r_{\mu\nu}=0$ for $\mu\ne\nu$). Here, we need only
\begin{align}
  \Gamma^r_{tt}=-\frac{g'_{tt}(r)}{2g_{rr}(r)}\>.
\end{align}
Since the $\dot{t}$ term of the Lagrangian has no
$\vartheta$ or $\varphi$ dependent prefactor, the equations for
$\vartheta$ and $\varphi$ just give $\Gamma^\vartheta_{tt}=
\Gamma^\varphi_{tt}=0$. So the only non-vanishing component of the
four-acceleration is in the $r$ direction. Moreover, from \eqref{eq:lagr_mparticle_spher_symm} we have
$g_{tt} c^2 \dot{t}^2=-c^2$, hence $\dot{t}^2=-1/g_{tt}$, and finally,
\begin{align}
  a^r =\frac{g'_{tt}(r)c^2}{2g_{tt}(r)g_{rr}(r)}\>.
\end{align}
This is the (outward directed) radial component of the
four-acceleration in the coordinate basis of the metric. The proper
acceleration $a$ is just the length of this vector, hence
\begin{align}
  a = \sqrt{a^\lambda a_\lambda} = \sqrt{g_{rr}\left(a^r\right)^2}
  = \frac{g'_{tt}(r)c^2}{2g_{tt}(r)\sqrt{g_{rr}(r)}}\>.
  \label{eq:proper_accel}
\end{align}
What will be the acceleration ``seen'' by a distant observer,
corresponding to this?

Let us find out what parallel transport gives. Since the answer will depend
on the path along which we do the transport, we should choose a
plausible path to get a meaningful result. A choice that suggests
itself is transport along a radial geodesic. We know by symmetry that
radial straight lines can be particle trajectories, i.e., geodesics of
space-time. Transporting the vector $a^\lambda$ from $r=r_1$ to
$r=r_2$, we must then require the covariant derivative with respect to
$r$ to vanish along the path:
\begin{align}
  a^\lambda_{;r} = a^\lambda_{,r}+\Gamma^\lambda_{r\mu}a^\mu = 0\>.
\end{align}
For $\lambda=t,\vartheta,\varphi$, the $\Gamma^\lambda_{rr}$ are
zero, so we find $a^\lambda_{,r}=0$, if $a^\mu=0$
for $\mu\ne r$. Since these components were zero to begin with,
they remain zero along the path. The only component that may change is
$a^r$. Here, we obtain
\begin{align}
   a^r_{;r} = a^r_{,r}+\Gamma^r_{rr}a^r= \pabl{a^r}{r} + \frac{g'_{rr}}{2 g_{rr}} a^r =0\>.
\end{align}
This first-order differential equation can be integrated using the
initial condition at $r_1$, which yields
\begin{align}
  a^r(r) = \frac{g'_{tt}(r_1) c^2}{2\sqrt{g_{rr}(r) g_{rr}(r_1)}\> g_{tt}(r_1)}\>,
\end{align}
and the length of this vector is
\begin{align}
  a(r) = \frac{g'_{tt}(r_1)c^2}{2g_{tt}(r_1)\sqrt{g_{rr}(r_1)}}\>,
  \label{eq:dist_proper_accel}
\end{align}
which is independent of $r$. Note that this is precisely the same
result as \eqref{eq:proper_accel}, because the radial coordinate $r$
in that equation \emph{is} the initial value $r_1$. So the proper
acceleration of the observer at $r_1$ is the same for observers at
any intermediate $r$ between $r_1$ and $r_2$ and at $r_2$
itself. In the end, this is not surprising: The proper
acceleration is a relativistic invariant, so it must be the same for
all observers. However, this is not necessarily what we would like the
local acceleration to translate into for the distant observer. The
fact that an observer at larger $r$ sees activities happening at
smaller $r$ slowed down in time should show up in a smaller value of
an acceleration for the distant observer than for the local
one. Hence, we might consider options different from parallel
transport to give meaning to the notion of ``acceleration at a distance''.

One way to proceed is to start from an uncontroversial relationship
such as the frequency change of a photon emitted at $r_1$ with
frequency $\nu_1$ and received at $r_2$ with frequency
$\nu_2$,
\begin{align}
  \nu_2 = \frac{\sqrt{-g_{tt}(r_1)}}{\sqrt{-g_{tt}(r_2)}}\, \nu_1\>,
  \label{eq:photon_frequencies}
\end{align}
and to use well-devised thought experiments to develop a description
for the distant observer at $r_2$ that is consistent with the local
description at $r_1$. The standard interpretation of
eq.~\eqref{eq:photon_frequencies} is in terms of time dilation, i.e.,
we would consider it equivalent to eq.~\eqref{eq:comp_prop_times}. In
fact, some authors \cite{okun99} consider this the \emph{only}
legitimate interpretation, saying that it is not the frequency of the
photon that changes but the perception of the frequency via measuring
devices that are subject to different rates of proper time. The
alternative view that the photon energy and, hence, its frequency,
decreases as the photon rises in a potential well is considered
inacceptable by these authors. However, it is an incontrovertible fact
that an electromagnetic wave emitted with frequency $\nu_1$ at $r_1$,
will be found to have frequency $\nu_2$ at $r_2$ by any experiment
capable of determining these frequencies. So to discuss the frequency shift
as something that did not ``really'' happen, seems moot.

What should be realized here is that the frequency is \emph{not} a
property of the photon \emph{alone}. Frequency is number of oscillations per
time. But the photon does not \emph{have} a time. The time in the
definition of frequency is an observer time. So the photon frequency
is a property of the relationship between the photon and the
observer. It is then certainly o.k.~to say that during the free fall
of the photon from $r_1$ to $r_2$, its internal properties do not
change. However, its frequency is determined not by its internal properties
alone but also by the frame of reference of the observer. In fact, the
same monochromatic electromagnetic wave will be observed to have
different frequencies by two observers at the same place,
if these have different velocities (Doppler effect), even though the
photons making up the wave are identical due to its monochromatic
nature. Just as frequency, energy is also a frame dependent
quantity.\footnote{This is particularly evident for kinetic energy,
  because the velocities that enter its definition change with a
  change of the frame of reference.}  Therefore, the point of view
seems legitimate that the energy change of the photon as it moves in a
gravitational field is due to the continually changing frame of
reference of local observers along its path. Since there exist frames,
in which the energy of the photon remains constant (e.g., the frame
with the global time, because that time is homogeneous), it is
possible to define a potential that does the book keeping vis-a-vis
such a frame. Via this potential, energy conservation can also be
implemented in the sequence of local frames passed by the
photon.\footnote{This would then make the aforementioned energy-loss
interpretation an acceptable alternative to the interpretation only in
terms of time dilation.}

Once we have established a relationship for the transformation of
local frequencies to distant ones via
eq.~\eqref{eq:photon_frequencies}, it is easy to set up a
transformation rule for energies. Since photon energies differ from
their frequencies only by a constant factor ($h$), we have, for these energies:
\begin{align}
 E_2 = \frac{\sqrt{-g_{tt}(r_1)}}{\sqrt{-g_{tt}(r_2)}}\, E_1\>.
  \label{eq:energy_transform}
\end{align}
But the same relationship must hold for \emph{any} form of energy, not
just for photons, if we require conservation of energy on converting
one form of energy into another. Otherwise, a perpetuum mobile may be
constructed \cite{kassner15}. Indeed, suppose that there are two forms
of energy, labeled by subscripts $a$ and $b$, for which the conversion
factors between local and distant energies would be different. Let the
total energy be $E_1=E_{a1}+E_{b1}$ in the local system at $r_1$ and
assume that
$E_2 =  E_{a1}/\varsigma_a +  E_{b1}/\varsigma_b$
with $\varsigma_b>\varsigma_a$. Convert the energy of type $b$
into type $a$ in the local system, so we obtain
$\tilde{E}_{a1} = E_{a1}+E_{b1}$ due to local energy
conservation. According to the distant observer, the energy is now
$\tilde{E}_2 =
\tilde{E}_{a1}/\varsigma_a =
E_2+E_{b1}\left({1}/{\varsigma_a}-{1}/{\varsigma_b}\right)
>E_2$, so for her, energy conservation does not hold. We may take
the view that the energy-containing system does not suffer any
internal changes when transported in free flight to $r_2$, so it will
arrive there with energy $\tilde{E}_2$. We convert this
completely to energy of type $b$. Then we have, due to local energy
conservation at $r_2$, $\hat{E}_b = \tilde{E}_2$ and this 
corresponds to an energy
$\hat{E}_1 =  \tilde{E}_2\varsigma_b=
(E_{a1} +E_{b1})\,{\varsigma_b}/{\varsigma_a}$.
Moving the energy on a free-fall trajectory back to
$r_1$ completes the cycle with 
$\hat{E}_1= E_1{\varsigma_b}/{\varsigma_a}>E_1$ at the disposal of
the observer at $r_1$. To exclude such a possibility, we must assume
\eqref{eq:energy_transform} to hold for any kind of energy.

\begin{figure}[!htb]
  \centering
  \includegraphics[height=6.0cm]{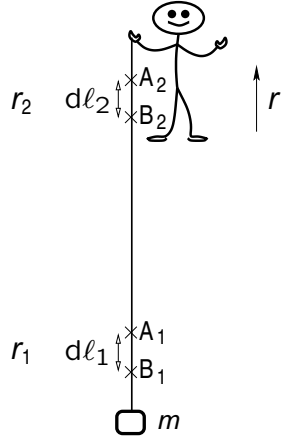}
  \caption{Tether used for an operational definition of forces and
    lengths at a distance. \label{fig:tether}}
\end{figure}

The tether then is a means to establish a relationship between local
and distant forces using the energy relationship, with the basic
arrangement shown in fig.~\ref{fig:tether}. A mass $m$ is suspended
from a massless inextensible tether and we ask what force has to be
exerted at its top to balance the gravitational force pulling on the
mass at its bottom.
Neither masslessness nor inextensibility are achievable in a strict
sense, but they can be attained with increasing accuracy at the price
of increasing effort (taking lighter and stronger materials). We can
do the experiment quasistatically, so special relativistic length
contraction will not play any role. Once the tether relationships are
confirmed over some finite distance experimentally, they can be
extended to arbitrary distances via the thought experiment. Due to the
inextensibility of the tether, if a proper length element $\D \ell$ is
spooled out at its top, its bottom end will go down by an equal proper
length element. The work done by the sinking mass is
$-\D E = F \D \ell$,\footnote{Force and proper length element carry a
  sign here. The positive direction is given by $\vec{e}_r$, so a
  negative $F$ is pointing ``downward'' and $\D\ell$ has the same sign
  as $\D r$.  } so the (radial) force may be calculated from the
energy via $F = - \D{E}/{\D\ell}$. Since $\D \ell$ is a proper length
element and hence invariant under changes of the frame of reference,
the transformation law for forces must be the same as that for
energies:
\begin{align}
 F_2 = \frac{\sqrt{-g_{tt}(r_1)}}{\sqrt{-g_{tt}(r_2)}}\, F_1\>.
  \label{eq:force_transform}
\end{align}
In fact, the tether relationship allows us to fully calculate the
radial local force $F(r)$ in terms of the metric coefficients, because
we know that the energy of a mass at rest is $E=mc^2$ in its local
frame. Setting $r_1=r$, we have, from \eqref{eq:energy_transform}
\begin{align}
  E_2(r_2,r) = \frac{\sqrt{-g_{tt}(r)}}{\sqrt{-g_{tt}(r_2)}}\,m c^2\>,
\end{align}
hence
\begin{align}
  F_2(r_2,r)&= -\abl{E_2}{\ell} = -\abl{E_2}{r}\abl{r}{\ell}  =  \frac{-1}{\sqrt{-g_{tt}(r_2)}}\frac{-g'_{tt}(r)}{2\sqrt{-g_{tt}(r)}} m c^2 \frac{1}{\sqrt{g_{rr}(r)}}\>.
              \label{eq:force_2}
\end{align}
From this, we may obtain the local force by use of \eqref{eq:force_transform}
\begin{align}
  F_1(r_1) &= \frac{\sqrt{-g_{tt}(r_2)}}{\sqrt{-g_{tt}(r_1)}}  F_2(r_2,r_1)
             = -m \frac{g'_{tt}(r_1) c^ 2}{2 g_{tt}(r_1)\sqrt{g_{rr}(r_1)}}
\end{align}
and this is just the mass times minus the proper acceleration from
\eqref{eq:dist_proper_accel}, that is, in the local frame, Newton's
law holds with the local acceleration given by the negative proper
acceleration. This corresponds to expectations; the force equation
\eqref{eq:force_transform} gives the correct answer for the local
result. We have consistency so far.

\begin{figure}[!htb]
  \centering
  \includegraphics[height=6.0cm]{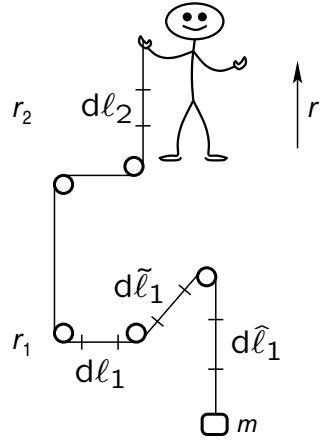}
  \caption{Spooling out the tether by $\D\ell_2$ at $r_2$ moves its
    bottom pieces by
    $\D\ell_1=\D\tilde{\ell}_1 =\D\hat{\ell}_1=\D\ell_2$ in the
    horizontal, oblique and vertical
    directions. \label{fig:contraption}}
\end{figure}

Shuler actually uses eq.~\eqref{eq:force_2}, with $r_2=\infty$ and the
abbreviation $g=\abs{g'_{tt}} c^2/2$ as the force (per unit mass)
countering the accelerating force from the point of view of a distant
observer. This would suggest consistency with the development
presented here. However, the derivation of eq.~\eqref{eq:force_2} is
not based on Shuler's approach to lengths at a distance. The tether
argument suggests
\begin{align}
  \D \ell_1 = \D \ell_2   \qquad \Leftrightarrow \qquad \sqrt{g_{rr}(r_1)} \,\D r_1
  =  \sqrt{g_{rr}(r_2)} \,\D r_2\>,
  \label{eq:length_transp_tether}
\end{align}
which for $r_2\to\infty$ reduces to $\D r_2=\D \ell_1$.  This can be
readily seen from fig.~\ref{fig:tether}. The crosses at $A_2$ and $A_1$
denote points marked on the tether before spooling out a length
increment $\D \ell$. After spooling, $A_2$ has moved to $B_2$ and
$A_1$ to $B_1$, and the proper distance traveled by the point $A_2$
that a nearby coordinate stationary observer at $r_2$ assigns to this
procedure is $\D \ell_2 = \D \ell$, that assigned by a similar
observer near $A_1$ to the motion of that point $\D \ell_1= \D
\ell$. This gives eq.~\eqref{eq:length_transp_tether}, so the tether
in fact is a tool for comparing lengths at a distance, too. Note,
moreover, that it can easily be extended to a comparator for
arbitrarily oriented length elements. This is demonstrated by the
contraption of fig.~\ref{fig:contraption}, where a few frictionless
pulleys have been added to show that length elements at a distance are
always directly related by equations without any factors $\rho_{rr}$,
when comparison is effected via a tether.

\subsection{An inconsistency}

This then suggests that Shuler's scaling relations between distant
lengths and his force calculation are mutually inconsistent. It is
easy to show that this is indeed the case. Shuler states the following
transformation rules for radial velocities, accelerations and forces
[his equations (3) and (5), with slightly abbreviated notation]
\begin{align}
  \begin{aligned}
    v_{2\text{rad}} &= v_{1\text{rad}} \,\rho_{rr}/\varsigma \\
    a_{2\text{rad}} &= a_{1\text{rad}} \,\rho_{rr}/\varsigma^2 \\
     F_{2\text{rad}} &= F_{1\text{rad}}/\varsigma\>,
\end{aligned}
\end{align}
and he obviously also agrees with
$E_{2\text{rad}} = E_{1\text{rad}}/\varsigma$, which follows, via the
photon frequency argument, from his $\D t = \varsigma \D\tau$. Since
this must also hold for kinetic energies and we are entitled to
consider slowly moving particles, for which the kinetic energy is
given by $m_{\text{in}} v^2/2$ we can derive a scaling law for the inertial mass $m_{\text{in}}$ in this relationship:
\begin{align}
  E_{2\text{kin}} =\frac12 m_{2\text{in}} v^2_{2\text{rad}} = \frac12 m_{2\text{in}}v^2_{1\text{rad}} \rho^2_{rr}/\varsigma^2  
                  &=  E_{1\text{kin}}/\varsigma =  \frac12 m_{1\text{in}}v^2_{1\text{rad}} /\varsigma \nonumber
  \\
  \Rightarrow\quad m_{2\text{in}}&=  m_{1\text{in}} \frac{\varsigma}{\rho_{rr}^2}\>.
                                   \label{eq:mass_conversion_1}
\end{align}
This says that the observed radial inertial mass at infinity is
reduced by a factor $\varsigma/\rho_{rr}^2$ with respect to its local
value. (The factor $\rho_{rr}$ is missing for the transverse
inertial masses, so the inertial mass will acquire a tensorial character,
to which no \emph{a priori} objections are made.)

However, we may also derive the relationship between local and distant
inertial masses from the force and acceleration equations
\begin{align}
  F_{2\text{rad}} = m_{2\text{in}} a_{2\text{rad}} = m_{2\text{in}}  a_{1\text{rad}} \rho_{rr}/\varsigma^2 
  &=  F_{1\text{rad}}/\varsigma =  m_{1\text{in}} a_{1\text{rad}}/\varsigma \nonumber
  \\
  \Rightarrow\quad m_{2\text{in}}&=  m_{1\text{in}}  \frac{\varsigma}{\rho_{rr}}\>,
            \label{eq:mass_conversion_2}                        
\end{align}
which gives a \emph{different} factor now between the two masses. The
results \eqref{eq:mass_conversion_1} and \eqref{eq:mass_conversion_2}
are incompatible with each other, unless $\rho_{rr}=1$. But Shuler
uses $\rho_{rr}$ values different from one in the rest of his
paper. His reasoning is based on inconsistent equations.


\subsection{Field strength measurement}

We are now in a position to discuss consistency of field strength
measurement. It will be useful to first verify what this means in
Newtonian physics. The force ``measured'' by a tether at its upper
end, $r_2$, when a mass was fastened to its lower end, $r_1$, is the
sum of the weight of the mass at $r_1$ and the weights of the length
elements of the tether at their respective positions in the
gravitational field.  But we required the tether to be massless, so
the total force will be due to the weight of the mass at the bottom
only. The tension in the tether will be a constant and so will be the
force measured by it, as long as the position $r_1$ is not
changed. Now let an observer at an intermediate position $r$
($r_1<r<r_2$) pick up the tether with a force just sufficient to
counterbalance the weight at the bottom. The tether above $r$ will go
slack and can even be cut off without destroying mechanical
equilibrium. CFS then shows up in the force to be applied by the
observer at $r$ being constant, i.e., independent of $r$, and equal to
the weight of the mass at $r_1$.

Consider the same situation in the spherically symmetric metric
\eqref{eq:spherically_symm_metric}. We cannot now assume that the
local tension in the tether is constant, because that will be
prevented by time dilation. The force that an observer at $r$ must
balance to pick up the weight is obviously given by
eq.~\eqref{eq:force_transform} with $r_2$ replaced by $r$:
\begin{align}
 F(r) = \frac{\sqrt{-g_{tt}(r_1)}}{\sqrt{-g_{tt}(r)}}\, F_1\>.
  \label{eq:force_intermed}
\end{align}
This is a local force at $r$ and if $\abs{g_{tt}(r)}$ decreases with
decreasing $r$, the magnitude of this force will increase towards
the bottom of the tether, an effect that is due to time dilation. What
will the value of this force be in the frame of the observer at $r_2$,
according to the rules of force transmission via tethers? We just have
to apply the same formula, but now between $r$ and $r_2$
\begin{align}
  \tilde{F}(r_2,r) &= \frac{\sqrt{-g_{tt}(r)}}{\sqrt{-g_{tt}(r_2)}}\,F(r)
                     =\frac{\sqrt{-g_{tt}(r_1)}}{\sqrt{-g_{tt}(r_2)}}\, F_1
  = F_2(r_2,r_1)\>.
\end{align}
For the observer at $r_2$, the force does not depend on $r$, it
appears constant along the tether and it agrees with the weight he
assigned to the mass before [via
eq.~\eqref{eq:force_transform}].\footnote{Moreover, he might assign a
  constant tension to the tether, a somewhat dubious
  procedure. Whether the tether breaks or not depends only on the
  local string tension, not on the one assigned from a distance, which
  can be much lower.  }

Hence, CFS is satisfied the same way it is in Newtonian physics,
regardless of the values taken by the metric coefficients $g_{tt}(r)$
and $g_{rr}(r)$.

A brief discussion of what is wrong with Shuler's argument may be in
order. He sets $F=m g /(\rho_{tt}\rho_{rr})$ (where $g$ is
$\abs{g'_{tt}(r)} c^2/2$) and interprets this $g$ as the acceleration
seen by the distant observer. Shuler then simply requires
$\rho_{tt}\rho_{rr}=\text{const}$ to keep the formula $F\propto m
g$. If he had correctly taken into account the scaling of inertial
mass obtained from his force relationship according to
\eqref{eq:mass_conversion_2}, i.e.,
$m_{\text{in}} = m/(\rho_{tt}\rho_{rr})$ and $F=m_{\text{in}} g$, he
would have obtained \emph{no} condition on the metric
coefficients. Another flaw of his argument is that the variations of
$r$ considered in his discussion are actually variations of the
position of the mass, i.e., of $r_1$, rather than variations of an
observer position, as they should. It certainly cannot be assumed that
the force in the tether does not change when the mass moves farther
into, or out of, the inhomogeneous gravitational field.

\subsection{Conservation of space-time and $g_{tt}g_{rr}=-1$}

Shuler then goes on to state that the relationship $g_{tt}g_{rr}=-1$
``for a century [...] has been thought to be an artifact of coordinate
choice'' \cite{shuler18}. He neglects to mention that indeed there are
mathematical proofs that this relationship is a consequence of
coordinate choice, such as the one given in
Ref.~\cite{jacobson07}. To portray it as a mere belief is a
misrepresentation. What is more, his argument being faulty, he himself
fails to prove anything about the structure of space-time.

In fact, it is difficult to understand why Shuler did not check his
argument by applying it to different metrics such as the Rindler
one
  \begin{align}
  \D s^2 = -\frac{\bar{g}^2 x^2}{c^4} \,c^2 \D t^2 + \, \D x^2 + \D y^2 + \D
  z^2\>,
  \label{eq:rindler}
\end{align}
describing a parallel gravitational field.\footnote{$\bar{g}$ is the
  proper acceleration of the observer at $x=c^2/\bar{g}$.} The tether
argument is known to work there as well \cite{kassner17a}. If it
provides a restriction on the product of a temporal and a spatial
coefficient of the metric because of an underlying structural 
feature of space-time, we should also obtain a restriction between the
coefficients $g_{tt}$ and $g_{xx}$ here. In
the example, this cannot be discussed away by declaring the
coordinates artificial.  The spatial coordinates are Cartesian and
directly related to length measurements in the Rindler frame. The time
coordinate is Einstein synchronized just as in the Schwarzschild
metric. Nevertheless, we have no constancy of the product
$g_{tt}g_{xx}g_{yy}g_{zz}$ which Shuler's arguments suggest to be a
property for any set of ``physical'' coordinates. Hence, Shuler's
approach falls to the same counterexample as those of Schiff and
Tangherlini.

Moreover, the way Shuler discusses his ``conservation principle'' of
space-time as derived from CPM and CFS, suggests that he does not seem
to be clearly aware of the difference between a conservation law and
an invariance principle. Invariance of space-time under Lorentz
transformations makes sure that its four-dimensional volume element
does not change due to length contraction or time dilation. But
space-time being an objective entity in geometrical theories, its
volume element is even invariant under \emph{arbitrary} coordinate
transformations. This does not, however, restrict the metric
coefficients. It just requires them to appear in the correct way in
the representation of a 4D volume element $\D\mathcal{V}$, i.e., we
have
\begin{align}
  \D\mathcal{V} = \sqrt{\abs{\det(g_{\mu\nu})}}\,\D^4 x
\end{align}
in arbitrary coordinates $x^\mu$.

To complete the discussion of the reasons for the inverse relationship
between $g_{tt}$ and $g_{rr}$ let us consider the equations for a
radial null geodesic in the metric
\eqref{eq:spherically_symm_metric}. We may obtain these from a
Lagrangian constructed from the metric with the help of an affine
parameter instead of the proper time (which is not defined for
light). Call the affine parameter $\omega$ and assume the geodesic to
be radial, so ${\D\vartheta}/{\D\omega}={\D\varphi}/{\D\omega}=0$ and
the Lagrangian reduces to
\begin{align}
  L = \left(\abl{s}{\omega}\right)^2 = g_{tt} c^2\dot{t}^2 + g_{rr}\dot{r}^2= 0\>,
\end{align}  
where now the dot signifies a derivative with respect to
$\omega$. Since $t$ is a cyclic coordinate, we have energy
conservation
\begin{align}
   g_{tt} \dot{t} = A = \text{const}
\end{align}
and instead of the equation of motion for $r$, we use the line element
itself to describe the dynamics of $r$:
\begin{align}
  g_{tt} c^2\dot{t}^2 + g_{rr}\dot{r}^2 &=0
  \quad \Rightarrow \quad \frac{A^2c^2}{g_{tt}} = - g_{rr}\dot{r}^2 \nonumber \\
  \dot{r}^2 &= -  \frac{A^2c^2}{g_{tt}g_{rr}}\>.
\end{align}
We see that constancy of the denominator on the right-hand side is
equivalent to $r$ being a linear function of $\omega$:
$r=B\omega+C \>\Rightarrow\> \dot{r}=B= \text{const}.$ This proves the
claim that for a metric of the form \eqref{eq:spherically_symm_metric}
$g_{tt} g_{rr}$ is a constant (which must be -1, if the metric becomes
Minkowskian for $r\to\infty$), iff $r$ is an affine parameter on
radial null geodesics itself (which means it must be a linear function
of any other affine parameter). Clearly, this is a \emph{coordinate
condition}.

$x$ in the Rindler metric is not an affine parameter and therefore, we
do not have $g_{tt} g_{xx}=\text{const}$. However, introducing
the alternative coordinate $\xi=({\bar{g}}/{2c^2})\, x^2$, we can
rewrite the Rindler metric as
\begin{align}
  \D s^2 = -\frac{2\bar{g} \xi}{c^2} \,c^2 \D t^2 + \frac{c^2}{2\bar{g} \xi}\, \D \xi^2 + \D y^2 + \D z^2\>,
  \label{eq:rindler2}
\end{align}
and herein, $\xi$ is of course an affine parameter on null geodesics
in the $x$ direction. However, $\xi$ is not a coordinate having a
direct interpretation in terms of length measurements by Rindler
observers.

\section{Generalization of Newton's universal law of gravitation and its consequences}
\label{sec:gen_newton_grav}

Up to this point, my criticism of Shuler's article was rather
destructive. There is little merit in this kind of endeavor although
it is necessary, if science is not to lose its reputation as a
self-correcting enterprise. A mere comment might have been in order to
just point out the errors in Shuler's paper. But then the length
restrictions of comments make it difficult to achieve the clarity
needed in refuting imprecise notions or inaccurate chains of
reasoning. Still, it might not have been worth bothering with, unless
there was the possibility to present a constructive aspect, too. This
is what I would like to do now.

Shuler did not give any good reasons for why we should have
$g_{tt}g_{rr}=-1$, but we know of course that it is true for the final
result. Let us now pretend that sufficient arguments for this point
have been given and the metric has been reduced to depending on a
single radial function. Shuler then uses various different assumptions
to construct a family of metrics, in which $g_{tt}$ takes the
particular forms
\begin{align}
  g_{tt} = -\left(1+n \frac{G M}{r c^2}\right)^{-2/n}\>,
\end{align}
with $n=-2$ (the Schwarzschild solution), $n=-1$ or $n=1$. The $n=1$
solution does not have an event horizon.  Shuler suggests to consider
certain strong-field experiments to distinguish between them
experimentally. Instead, I will give a theoretical argument strongly
favoring one of these solutions.

The basic idea is to find an appropriate
relativistic generalization of Newton's universal law of gravitation
and then to use this to restrict the
metric. Unfortunately, the best-known form of Newton's law
\begin{align}
  \boldsymbol{F}_g &= - G\, \frac{m M}{r^2} \,\boldsymbol{e}_r
                     \label{eq:newton_force}                     
\end{align}
gives the gravitational force via an action at a distance (even though
it can be reinterpreted in local terms and Newton was convinced that
such an interpretation is the only sensible one). One might try to
gain intuition from Maxwell's equations generalizing Coulomb's law to
the relativistic case, but to start from \eqref{eq:newton_force} does
not look too promising. Moreover, the relativistic transformation laws
for forces are not trivial, which is a second obstacle.

The standard \emph{local} form of the law, 
\begin{align}
  \Delta \Phi_g(\vec{r}) &= 4\pi G \rho(\vec{r})\>,
         \label{eq:newton_poisson}                   
\end{align}
a Poisson equation with the mass density $\rho(\vec{r})$ as source
term, introduces a potential for the force, but a simple and
straightforward generalization does not suggest itself, the equation
being static and potentials only slightly less difficult than
forces. Moreover, with hindsight we know that general relativity has
more than one potential (each element of the metric may be considered
one) and it is not clear how to go from one to several potentials in a
compelling manner.

A variant of \eqref{eq:newton_poisson} that is physically much more transparent reads 
\begin{align}
  \Phi_g(\vec{r_c}) &= \overline{\Phi}_g\vert_{\left\{\vec{r}\left|\right. \,\abs{\vec{r}-\vec{r_c}}=\varepsilon\right\}} -\frac{G}{2\varepsilon} m(\vec{r_c},\varepsilon)\>.
\label{eq:newton_feynman}
\end{align}
It gives the potential at the center $\vec{r_c}$ of a small ball with
radius $\varepsilon$ as the \emph{average} over the potential on the
\emph{surface} of the ball and a correction term containing the
\emph{gravitating mass in the ball}. Of course,
eq.~\eqref{eq:newton_feynman} is just an integrated version of
\eqref{eq:newton_poisson}, obtained with the help of the free-space
Green's function, but it is much easier to interpret than
\eqref{eq:newton_poisson}. It tells us that when there is no mass
density around the point considered ($\vec{r_c}$), then the
gravitational potential, if it is not constant in the ball, does not
have a minimum at $\vec{r_c}$, as there must be a smaller potential
somewhere on the surface of the ball. So no test particle will have a
stable position anywhere in vacuum. Second, if there is a mass
density, it will reduce the potential, so the gravitational force is
attractive. Actually, the variant \eqref{eq:newton_feynman} is given
by Feynman in a YouTube video \emph{Many Mathematical Representations
  and Resulting Paradoxes}. He argues that physicists ought to keep in
mind different mathematical representations of the same law of nature,
because they do not know ahead of time, which one is the best to
generalize when something breaks due to more precise experiments and a
new theory is needed.

Neither of the three representations of Newton's law is too
well-suited for generalization, so I will suggest a fourth, the goal
being a statement that not only describes gravitation at a single
point in space but provides information about tidal forces, which,
after all, are an essential feature of gravity. Let us introduce the
concept of a dust ball. This is a cloud of test particles,
i.e.,~particles so small that they do not disturb the gravitational
field and do not interact gravitationally with each other. Their
motion will of course be affected by an external gravitational
field. We will assume our dust balls to be small and initially
spherical or ellipsoidal, which means that after a sufficiently small
deformation they will still be ellipsoids.\footnote{This is true as
  long as deformations are small enough for changes of higher than
  quadratic order in their amplitudes to be negligible. The ellipsoid
  may then change the position of its center and its orientation as
  well as the lengths of its semi-major axes. But it will remain an
  ellipsoid.}  Then I claim that Newton's law can be formulated as
follows: \emph{Given a sufficiently small freely falling dust ball,
  the particles of which are initially at rest with respect to each
  other, the rate at which it will start to shrink is proportional to
  its volume times the mass density at its center.} Quantitatively:
\begin{align}
  \left.\frac{\ddot{V}}{V}\right\rvert_{t=0} = -4\pi G\rho(\vec{r})\>.
  \label{eq:dustball_newton}
\end{align}
The formulation is a minor modification of a similar law given by Baez
and Bunn in an article about the meaning of Einstein's
equation \cite{baez05}. A few comments may be useful. For convenience,
the initial time has been set equal to zero. Since the particles are
assumed to be initially not moving with respect to each other, the
volume rate of change is zero to linear order in $t$, so we need to
consider the quadratic order, leading to a second derivative in time
(a dot is a Newtonian time derivative). We could, instead of invoking
volume and density, simply  have taken $\ddot{V}$ to be proportional to
the mass inside the volume -- for small enough volumes and continuous
$\rho$, the two formulations are equivalent. The one given in
eq.~\eqref{eq:dustball_newton} has been chosen for convenience. 
An interesting feature is that
the equation contains neither forces nor potentials -- we describe
their effects by motion (presupposing, of course, Newton's second axiom
to hold). 

\begin{figure}[!htb]
  \centering
  \includegraphics[height=8.0cm]{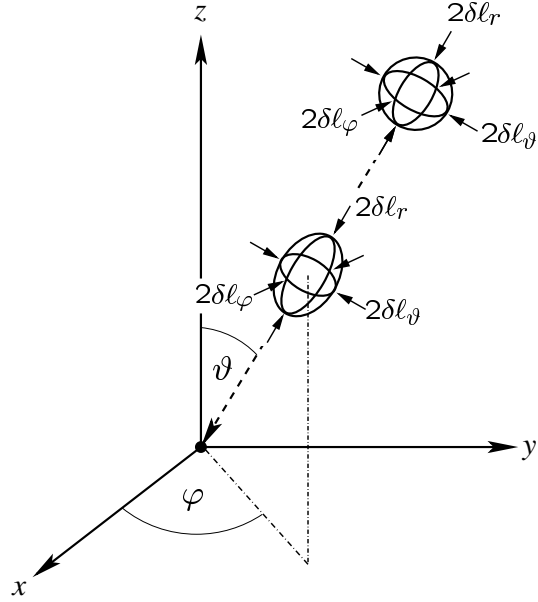}
  \caption{Dust ball falling radially towards a point mass at the
    origin and deforming in the process. \label{fig:dustball}}
\end{figure}

To show that eq.~\eqref{eq:dustball_newton} is equivalent to Newton's
law in its standard formulation \eqref{eq:newton_force}, consider a
gravitating point mass, towards which a dust ball is falling that was
initially at rest with respect to it, see fig.~\ref{fig:dustball}. 
We assume the point mass to be structureless, so its gravitational
field will have spherical symmetry, i.e., it will be a central force
field. Also, the mass is assumed to be constant, so the field will not
be time dependent. It is known that time and velocity independent
central forces are conservative, so the field has a potential
$\Phi(r)$.  The initial rest state means that each test particle is
falling on a radial trajectory towards the center. Call the semi-axes
of the ball along the directions of a spherical coordinate system
$\delta\ell_r$, $\delta\ell_\vartheta$, and $\delta\ell_\varphi$. For
symmetry reasons, the orientation of these axes will not change during
a short interval of fall. The volume of the ellipsoid is
$V=\frac{4\pi}{3} \delta\ell_r\delta\ell_\vartheta
\delta\ell_\varphi$, its rate of change may then be expressed as
\begin{align}
  \frac{\ddot{V}}{V}\Bigr\rvert_{t=0} &=    \frac{\delta\ddot{\ell}_r}{{\delta\ell}_r}+
                                        \frac{\delta\ddot{\ell}_\vartheta}{{\delta\ell}_\vartheta}  +
                                        \frac{\delta\ddot{\ell}_\varphi}{{\delta\ell}_\varphi} \>\>\Bigr\rvert_{t=0}\>,
                                        \label{eq:ddot_V_newton}
\end{align}
because first-order time derivatives of the $\delta\ell_\mu$ vanish. Writing
$\delta\ell_r=\delta r$, $\delta\ell_\vartheta= r \delta\vartheta$ und
$\delta\ell_\varphi= r \sin\vartheta\delta\varphi$, we obtain, taking
the time derivatives and dropping first order derivatives at the end
\begin{align}
\frac{\delta\ddot\ell_r }{\delta\ell_r} &= \frac{\delta \ddot r}{\delta r}\>,
\label{eq:ddotellr}
\\
\frac{\delta\ddot\ell_\vartheta }{\delta\ell_\vartheta} &=\frac{\ddot r}{r} 
+ \frac{\delta \ddot\vartheta}{\delta\vartheta}\>,
\label{eq:ddotellthet}
\\
  \frac{\delta\ddot\ell_\varphi }{\delta\ell_\varphi} &=\frac{\ddot r}{r} 
                                                      +  \cot\vartheta \,\ddot\vartheta
                                                      +  \frac{\delta \ddot\varphi}{\delta\varphi}\>.
\label{eq:ddotellphi}
\end{align}
Because particles fall along straight radial lines towards the center,
$\delta\vartheta$ and $\delta\varphi$ are time independent -- the
$\vartheta$ and $\varphi$ coordinates of a particle do not change
during fall. Therefore, the last two formulas will reduce to just
$\ddot{r}/r$ on their right-hand sides and we obtain for the rate of
volume change the simple result
\begin{align}
 \frac{\ddot{V}}{V}\Bigr\rvert_{t=0} &=\frac{\delta \ddot r}{\delta r} + 2\frac{\ddot r}{r}\>.
\end{align}
According to our dust ball form of Newton's law (DBNL), eq.~\eqref{eq:dustball_newton}, we must have
\begin{align}
  \frac{\delta \ddot r}{\delta r} + 2\frac{\ddot r}{r} = 0\>,
  \label{eq:dust_ball_point_vac}
\end{align}
because outside of the attracting point mass, the mass density is
zero. (The density of the test particles is negligible.) It is then
easy to express $\ddot r$ and $\delta \ddot r$ with the help of the potential (a prime denotes a derivative w.r.t. $r$):
\begin{align}
  \ddot r &= -\Phi'(r)\>, \qquad
 \delta \ddot  r = -\Phi'(r+\delta r)+ \Phi'(r) = -\Phi''(r) \delta r \>.
\end{align}
Substituting this into \eqref{eq:dust_ball_point_vac}, we find
\begin{align}
  -\Phi''(r)-\frac{2 \Phi'(r)}{r} = 0\>.
  \label{eq:lin_dgl_newt}
\end{align}
This is a linear differential equation for the potential, the general
solution of which is straightforward to find:
\begin{align}
  \Phi(r) = \frac{c_1}{r}+c_2\>.
\end{align}
The constant $c_2$ is additive and may be chosen equal to zero. To determine $c_1$, we write out the DBNL with a $\delta$ function mass density:
\begin{align}
  -\Phi''(r)-\frac{2 \Phi'(r)}{r} = -\Delta \Phi(r)= -4\pi G M \delta(\vec r)\>.
  \label{eq:DBNL_poisson}
\end{align}
Here, we have exploited that the differential operator in $r$ applied to $\Phi$ on the left-hand side is just the radial part of the Laplacian. We integrate over a small sphere with radius $\varepsilon >0$ about the origin and use Gauss's divergence theorem to find
\begin{align}
   \oint_{r=\varepsilon} \nabla \Phi \cdot\D \vec S
  &= \int_{r\le\varepsilon} 4\pi G M \delta(\vec r) \,\D^3 r = 4\pi G M\nonumber
  \\
  -\frac{c_1}{\varepsilon^2} \, 4\pi \varepsilon^2 &= 4\pi G M
  \qquad \Rightarrow\qquad
  c_1 = -GM \>.
        \label{eq:determine_c1}
\end{align}
Hence, the potential is $\Phi(r) = -{GM}/{r}$, leading to a force on a
test mass $m$ given by $\vec F = -G{mM}/{r^2}\,\vec e_r$. Since the
calculation works both ways, i.e., we may derive the standard form of
Newton's law from DBNL and we may start from the potential of Newton's
law and trace the calculation backwards from \eqref{eq:determine_c1}
to \eqref{eq:dustball_newton}, the equivalence of DBNL and Newton's
universal law of gravitation has been shown.  DBNL is just
another way to formulate Newton's gravitational law.

Generalization of \eqref{eq:dustball_newton} to a relativistic
equation is still not easy because of its right-hand side. In Newton's
theory, the field source is a scalar density. With hindsight, we know
that its generalization will lead to the divergence of a
four-tensor. However, our interest is to obtain results on a special
solution or class of solutions of the full theory without using knowledge from
the field equations (which would immediately allow us to conclude
what the generalization must look like). So we eschew attempts to
generalize DBNL in its full glory. But in vacuum the right-hand side
is zero, eq.~\eqref{eq:dustball_newton} looks like a purely
kinematic relation and we know how to deal with those.

What do we have to do? We replace the Newtonian absolute time by the
proper time of the test particles or more precisely by the proper time
of an appropriately chosen representative (the center particle of the
ball). Moreover, we require the law to hold in a local inertial
frame\footnote{Speaking of a local inertial frame in which tidal
  forces are being felt, is tricky.  Gravitation is treated like any
  other field quantity this way. The point will be reconsidered
  below.} rather than in the global Newtonian one, of which absolute
space is a representative. The \emph{relativistic dust ball law} for
motion in a gravitational field in vacuum (DBV) then takes the form
\begin{align}
  \left.\frac{\ddot{V}}{V}\right\rvert_{\tau=0} = 0\>,
  \label{eq:dustball_einstein_vac}
\end{align}
all quantities being evaluated in the comoving local frame
of the  dust ball center particle. Overdots
 now mean derivatives with respect to proper time again.

Let us now assume that the central prerequisite for the derivation of
Shuler's family of gravities, the relationship
$g_{tt}g_{rr}=\text{const}$ has found some sound justification or we
have simply been told that it is true. Can we then use
\eqref{eq:dustball_einstein_vac} to further constrain the metric?

In terms of the rates of changes of the semi-axes, the volume rate of
change is still given by \eqref{eq:ddot_V_newton}, with the time being
replaced by the proper time of the center particle. In the
metric \eqref{eq:spherically_symm_metric},
$\delta \ell_r= \sqrt{g_{rr}} \delta r$, so instead of
eq.~\eqref{eq:ddotellr}, we obtain
\begin{align}                                                                                                  \frac{\delta\ddot\ell_r }{\delta\ell_r}
  &= \frac{\delta \ddot r}{\delta r}+\frac{g'_{rr}}{2 g_{rr}}\ddot{r}\>.  \label{eq:ddotellr_metric}                                          
\end{align}
The equations for the rates of change of
$\delta\ell_\vartheta$ and $\delta\ell_\varphi$ keep the forms
\eqref{eq:ddotellthet} and \eqref{eq:ddotellphi}, for obvious
reasons. These are quantities calculated in the coordinate stationary
frame given by the metric. What we need, however, are the
corresponding quantities in the frame of the falling center particle
of the dust ball. At first sight, it might seem that these must be the
same, because that falling frame is at rest with respect to the
stationary frame at time $t=\tau=0$ and its velocity grows linearly in
$t$ or $\tau$, is therefore negligible as $\tau\to0$. However, we have
to evaluate a second derivative of a position-like quantity and if
that quantity grows as $\tau^2$, i.e., remains small for small $\tau$,
the second derivative with respect to $\tau$ will still give a finite
contribution at $\tau=0$. To evaluate this contribution, a local
Lorentz transformation from the coordinate stationary to the
momentarily comoving freely falling frame may be
performed \cite{kassner17b}. As it turns out, this leaves the formulas
for the polar and azimuthal semi-axes \eqref{eq:ddotellthet} and
\eqref{eq:ddotellphi} unchanged (the velocity of the center particle
does not change in time along these two directions) but modifies
\eqref{eq:ddotellr_metric} into
\begin{align}
  \frac{\delta\ddot\ell_{cr} }{\delta\ell_{cr}}
  &= \frac{\delta \ddot r}{\delta r}+\frac{g'_{rr}}{2 g_{rr}}\ddot{r}-\frac{g_{rr}\ddot r^2}{c^2}\>,
    \label{eq:ddotellr_center}                                          
\end{align}
where the additional subscript $c$ is a reminder that this is a
quantity referring to the local freely falling frame of the center
particle.

For symmetry reasons, $\vartheta$ and $\varphi$ are constant for each
falling particle. The equation determing $\ddot r$ now follows from
the Lagrangian \eqref{eq:lagr_mparticle_spher_symm}, where the last
two terms may again be omitted for radial geodesics.  $t$ is cyclic
and as an equation for $r$ we may take the definition of the
Lagrangian. This leads to
\begin{align}
  g_{tt} \dot{t} &= A = \text{const}\>, \nonumber \\
  g_{tt} c^2\dot{t}^2 + g_{rr}\dot{r}^2 &=-c^2 
  \qquad\Rightarrow\quad A^2 c^2 +  g_{tt} \left(g_{rr}\dot{r}^2+c^2\right) = 0\>.
\end{align}
Differentiating the last equation with respect to $\tau$, we get rid of the constant $A$ and find
\begin{align}
  \ddot r = -c^2\frac{g'_{tt}}{2g_{tt}g_{rr}}-\frac{\left(g_{tt}g_{rr}\right)'}{2g_{tt}g_{rr}} \dot r^2\>.
\end{align}
Using $g_{tt}g_{rr}=-1$, this simplifies enormously,
\begin{align}
  \ddot r &=       \frac{ c^2}{2} g'_{tt}
\qquad  \Rightarrow \qquad \delta \ddot r =  \frac{c^2}{2} g''_{tt}\delta r \>.
                                     \label{eq:delta_ddot_r_schw}
\end{align}
Another consequence of the value $-1$ of this product is that the
second and the third terms on the right-hand side of  
\eqref{eq:ddotellr_center} cancel each other, so we have
${\delta\ddot\ell_{cr} }/{\delta\ell_{cr}} = {\delta \ddot r}/{\delta
  r}$. Evaluating this together with  Eqs.~\eqref{eq:ddotellthet} and
\eqref{eq:ddotellphi} in terms of  \eqref {eq:delta_ddot_r_schw}
and inserting into the DBV \eqref{eq:dustball_einstein_vac}, we obtain
\begin{align}
  c^2\frac{g''_{tt}}{2}+c^2 \frac{g'_{tt}}{r} = 0\>,
\end{align}
which is basically the same equation as the one we had for the
potential $\Phi(r)$ [eq.~\eqref{eq:lin_dgl_newt}], hence
$g_{tt}(r)=c_1/r+c_2$. With the boundary condition
$\lim_{r\to\infty} g_{tt}(r) = -1$, we obtain $c_2 = -1$. To determine
the constant of integration $c_1$, we may refer to the first equation of
\eqref{eq:delta_ddot_r_schw} and require that this equation reduces to its
Newtonian limit for large $r$, hence
\begin{align}
  \ddot r &= -\frac{c^2}{2} \frac{c_1}{r^2} \sim - \frac{GM}{r^2} \qquad (r\to\infty)
\>,
\end{align}
leading to $c_1=2GM/c^2$ 
and
\begin{align}
  g_{tt}(r) &= -\left(1-\frac{2GM}{c^2 r}\right)\>,
  \qquad
  g_{rr}(r) =\left(1-\frac{2GM}{c^2 r}\right)^{-1}\>,
\end{align}
which is the exact result for the Schwarzschild metric. This is
therefore the only one out of Shuler's family of solutions that is
compatible with the generalization \eqref{eq:dustball_einstein_vac} of
Newton's gravitational law. Apparently, the others can be ruled out
without strong-field considerations.

It should be added that the assumption $g_{tt}g_{rr}=-1$ is
unnecessary. Rather, this relation can be \emph{derived} as part of
the solution using the DBV law \eqref{eq:dustball_einstein_vac}.  All
that has to be done is to consider a second (independent)
configuration of a dust ball freely falling in the metric, a circular
orbit suggesting itself as the simplest choice. The full calculation
has been presented in Ref.~\cite{kassner17b}. 

In Ref.~\cite{kassner17b}, the DBV law was given as a physical
statement of Einstein's field equations in vacuum, because Baez and
Bunn had shown it to be just this \cite{baez05}. Here, it is justified
as a postulate by pointing out that it is a relativistic
generalization of Newton's law in vacuum, obtained by the usual
heuristic rules. In fact, its statement reminds of an 
application of the equivalence principle to a Newtonian formulation,
valid in a comoving local frame. However, DBV goes beyond at
least the EEP. What the EEP says is that in describing local
\emph{non-gravitational} physics, we may find, given a certain level of
accuracy of our measuring devices, a freely falling comoving system in
which, if we choose its spatial dimensions sufficiently small and
restrict experiments to a sufficiently short duration, all
experimental results will be the same no matter what the velocity or
the spatiotemporal location of the system. The theory predicting these
results is special relativity, so the freely falling system is a local
inertial system.  But DBV talks about tidal forces which are not
predicted by special relativity. In a way, DBV pretends that gravitational
fields can be incorporated into the theory much the same way as other
fields, and are not different in this respect from, say,
electromagnetic fields, for which it is no problem to formulate a law
in a local inertial frame. But gravity, if detectable, destroys the
inertial nature of the local system... Alternatively (and preferably), we
could see DBV as transcending the limits of the notion of local \emph{inertial}
system. It tells us what we should expect to see, if we increase the
accuracy of our measuring devices, in what used to be a local inertial
frame, sufficiently to detect tidal forces, which makes the system
non-inertial. Instead, we could increase the size of the system (or
wait long enough) until we see tidal forces with the current accuracy
of our devices. It then appears that DBV is somewhat extending the
boundaries of the system to match the physics described by the
equivalence principle to the physics of the ``world at large''.

An interesting question to ponder is whether, given Newton's law as a
non-relativistic description of gravity, the DBV law is a requirement
of the \emph{strong} equivalence principle, stating that the results of
sufficiently local experiments, \emph{including gravitational ones}, do not
depend on the velocity and spatiotemporal location of the freely
falling frame in which they are performed \cite{gron07}. Tentatively,
this author would answer this in the affirmative.

Now, it is certainly conceivable that general relativity becomes
incorrect for sufficiently strong gravitational fields. The field
equations are derived from the Einstein-Hilbert action, in which the
Lagrangian density of the field is just the scalar curvature. If it were a
nonlinear function(al) of the scalar curvature instead, starting with
the same linear term, then weak-field predictions of the theory would
remain unchanged whereas the strong-field sector would look different.
It appears, however, that the form of DBV would not change, because
curvature effects can be made arbitrarily small by sufficient
reduction of the size of the local freely falling system (including
its temporal extension). This then suggests that the Schwarzschild
geometry would still be a solution of such a modified theory of
gravity.\footnote{Since the scalar curvature $R$ of the Schwarzschild
  solution is exactly zero everywhere outside the central singularity,
  it is perfectly possible for that solution to be the stationary
  point of other actions formulated in terms of the scalar curvature,
  besides the Einstein-Hilbert one. After all, nonlinear terms in $R$
  of these actions will become negligible both for the Schwarzschild
  solution and small variations about it.} Even when additional fields
enter, such as the scalar field in the Brans-Dicke theory
\cite{brans61}, this may remain true. Indeed, the Schwarzschild result
is an exact black-hole solution of the Brans-Dicke theory, with
constant scalar field. It just does not describe the field outside a
\emph{finite-sized} spherical mass distribution correctly anymore.
Generally speaking, it would seem that the study of stationary vacuum
solutions is not the best way to discriminate between different
strong-field variants of metric theories of gravity. Rather, alternate
theories should generate predictions for \emph{dynamic} solutions to
be tested experimentally.

While it is a positive feature of Shuler's article that it proposes
experiments by which to make a distinction between the members of his
family that all behave the same in the weak-field case, he does not
really do justice to the strong-field experiments that are already
available. Observations of the decrease of the orbital period of the
Hulse-Taylor pulsar \cite{weisberg10} cannot really be called
weak-field probes anymore, and they agree well with the predictions of
general relativity. The direct observation of gravitational waves
\cite{abbott16} definitely tests the strong-field domain of the theory
and it is somewhat moot to point out that inaccuracies of the
determination of mass parameters of the partners of a black-hole
merger still exceed 10\% typically. The wave form and phase of the
gravitational wave signal are highly specific and this in itself
constitutes a quantitative test of the theory at strong
fields. Obviously, Shuler's results cannot be readily tested against
this, because he does not have a field theory and cannot predict
dynamical situations in any detail. Still, it appears that one of the
members of his family (the ``proximity-gravity'' one, $n=1$) may be
immediately ruled out due to the presence of a ring-down signal with
an approximately constant frequency at the end of the gravitational wave
train \cite{abbott16}. This signal is a consequence of the
formation of a horizon of the final black hole that is quickly damped
to constant shape. Since no horizon would be formed in the
``proximity-gravity'' model, instead a continuing gravitational wave
signal from the final massive object (that is still moving and
producing a time dependent quadrupole moment) would have to be
expected. Other results than a final black hole (such as worm holes)
would lead to echoes in the gravitational wave signal. These are
actively being searched for. But most of the more than ten detection
events reported so far \cite{abbott18} are readily interpreted in
terms of a final black hole emerging, in quantitative agreement with
GR predictions.

\section{Conclusions}
\label{sec:conclusions}

In summary, it has been shown that local energy and momentum
conservation are an integral feature of the dynamic equations for
particles from any metric theory of gravity. Therefore, momentum
conservation considerations are not helpful in replacing the field
equations when it comes to restricting the metric to a physically
possible solution. This is at least true, if the conservation law does
not refer to the momentum of the gravitational field itself (i.e.,
momentum carried by gravitational waves).

Consistency of field strength measurement is another requirement that
is automatically satisfied by any metric, provided measurements at a
distance are interpreted in a consistent way. Shuler's manner of
establishing force measurements at a distance corresponds to the
tether approach discussed, whereas his connection of distant with
local velocities corresponds to a different convention of comparing
lengths at a distance, incompatible with the force definition
established by tethers.


No physical fact can be identified corresponding to Shuler's
``conservation of space-time''. Invariance of space-time volume
elements is of course satisfied in GR and other metric theories.

The exact Schwarzschild metric may be derived using the EEP, SR, NL
and a generalization of Newton's universal law of gravitation in
vacuum to a general relativistic setting by standard rules applicable
to kinematic laws, i.e., the replacement of time by a local proper
time and the replacement of Newton's absolute space frame by a local
freely falling frame. Only one member of Shuler's family of gravities
satisfies the constraints on the metric following from this approach,
so the others should be discarded. The surviving gravitational theory
is just ordinary GR.


\begin{thebibliography}{10}%
\makeatletter
\providecommand \@ifxundefined [1]{%
 \ifx #1\undefined \expandafter \@firstoftwo
 \else \expandafter \@secondoftwo
\fi
}%
\providecommand \@ifnum [1]{%
 \ifnum #1\expandafter \@firstoftwo
 \else \expandafter \@secondoftwo
\fi
}%
\providecommand \enquote [1]{``#1''}%
\providecommand \bibnamefont  [1]{#1}%
\providecommand \bibfnamefont [1]{#1}%
\providecommand \citenamefont [1]{#1}%
\providecommand\href[0]{\@sanitize\@href}%
\providecommand\@href[1]{\endgroup\@@startlink{#1}\endgroup\@@href}%
\providecommand\@@href[1]{#1\@@endlink}%
\providecommand \@sanitize [0]{\begingroup\catcode`\&12\catcode`\#12\relax}%
\@ifxundefined \pdfoutput {\@firstoftwo}{%
 \@ifnum{\z@=\pdfoutput}{\@firstoftwo}{\@secondoftwo}%
}{%
 \providecommand\@@startlink[1]{\leavevmode}%
 \providecommand\@@endlink[0]{}%
}{%
 \providecommand\@@startlink[1]{%
  \leavevmode
  \pdfstartlink
   attr{/Border[0 0 1 ]/H/I/C[0 1 1]}%
   user{/Subtype/Link/A<</Type/Action/S/URI/URI(#1)>>}%
  \relax
 }%
 \providecommand\@@endlink[0]{\pdfendlink}%
}%
\providecommand \url  [0]{\begingroup\@sanitize \@url }%
\providecommand \@url [1]{\endgroup\@href {#1}{\urlprefix}}%
\providecommand \urlprefix [0]{URL }%
\providecommand \Eprint[0]{\href }%
\@ifxundefined \urlstyle {%
  \providecommand \doi [1]{doi:\discretionary{}{}{}#1}%
}{%
  \providecommand \doi [0]{doi:\discretionary{}{}{}\begingroup
  \urlstyle{rm}\Url }%
}%
\providecommand \doibase [0]{http://dx.doi.org/}%
\providecommand \Doi[1]{\href{\doibase#1}}%
\providecommand \bibAnnote [3]{%
  \BibitemShut{#1}%
  \begin{quotation}\noindent
    \textsc{Key:}\ #2\\\textsc{Annotation:}\ #3%
  \end{quotation}%
}%
\providecommand \bibAnnoteFile [2]{%
  \IfFileExists{#2}{\bibAnnote {#1} {#2} {\input{#2}}}{}%
}%
\providecommand \typeout [0]{\immediate \write \m@ne }%
\providecommand \selectlanguage [0]{\@gobble}%
\providecommand \bibinfo [0]{\@secondoftwo}%
\providecommand \bibfield [0]{\@secondoftwo}%
\providecommand \translation [1]{[#1]}%
\providecommand \BibitemOpen[0]{}%
\providecommand \bibitemStop [0]{}%
\providecommand \bibitemNoStop [0]{.\EOS\space}%
\providecommand \EOS [0]{\spacefactor3000\relax}%
\providecommand \BibitemShut [1]{\csname bibitem#1\endcsname}%
\bibitem{weinstein18}%
  \BibitemOpen
  \bibfield{author}{%
  \bibinfo {author} {\bibfnamefont{G.}~\bibnamefont{Weinstein}},\ }%
  \bibfield{title}{%
  \enquote{\bibinfo {title} {{Why did Einstein reject the November tensor in
  1912–1913, only to come back to it in November 1915?}}.}\ }%
  \bibfield{journal}{%
  \bibinfo {journal} {Stud. Hist. Phil. Sci. B}\ }%
  \textbf{\bibinfo {volume} {62}},\ \bibinfo {pages} {98--122} (\bibinfo {year}
  {2018})%
  \bibAnnoteFile{NoStop}{weinstein18}%
\bibitem{kassner15}%
  \BibitemOpen
  \bibfield{author}{%
  \bibinfo {author} {\bibfnamefont{K.}~\bibnamefont{Kassner}},\ }%
  \bibfield{title}{%
  \enquote{\bibinfo {title} {{Classroom reconstruction of the Schwarz\-schild
  metric}},}\ }%
  \bibfield{journal}{%
  \bibinfo {journal} {Eur. J. Phys.}\ }%
  \textbf{\bibinfo {volume} {36}},\ \bibinfo {pages} {065031 (1--20),}
  (\bibinfo {year} {2015})%
  \bibAnnoteFile{NoStop}{kassner15}%
\bibitem{kassner17a}%
  \BibitemOpen
  \bibfield{author}{%
  \bibinfo {author} {\bibfnamefont{K.}~\bibnamefont{Kassner}},\ }%
  \bibfield{title}{%
  \enquote{\bibinfo {title} {{A physics-first approach to the Schwarzschild
  metric}},}\ }%
  \bibfield{journal}{%
  \bibinfo {journal} {Adv. Stud. Theor. Phys.}\ }%
  \textbf{\bibinfo {volume} {11}},\ \bibinfo {pages} {179--212} (\bibinfo
  {year} {2017})%
  \bibAnnoteFile{NoStop}{kassner17a}%
\bibitem{kassner17b}%
  \BibitemOpen
  \bibfield{author}{%
  \bibinfo {author} {\bibfnamefont{K.}~\bibnamefont{Kassner}},\ }%
  \bibfield{title}{%
  \enquote{\bibinfo {title} {{Dustball physics and the Schwarzschild
  metric}},}\ }%
  \bibfield{journal}{%
  \bibinfo {journal} {Am. J. Phys.}\ }%
  \textbf{\bibinfo {volume} {85}},\ \bibinfo {pages} {619--627} (\bibinfo
  {year} {2017})%
  \bibAnnoteFile{NoStop}{kassner17b}%
\bibitem{gruber88}%
  \BibitemOpen
  \bibfield{author}{%
  \bibinfo {author} {\bibfnamefont{R.~P.}\ \bibnamefont{Gruber}}, \bibinfo
  {author} {\bibfnamefont{R.~H.}\ \bibnamefont{Price}}, \bibinfo {author}
  {\bibfnamefont{S.~M.}\ \bibnamefont{Matthews}}, \bibinfo {author}
  {\bibfnamefont{W.~R.}\ \bibnamefont{Cordwell}},\ and\ \bibinfo {author}
  {\bibfnamefont{L.~F.}\ \bibnamefont{Wagner}},\ }%
  \bibfield{title}{%
  \enquote{\bibinfo {title} {{The impossibility of a simple derivation of the
  Schwarzschild metric}},}\ }%
  \bibfield{journal}{%
  \bibinfo {journal} {Am. J. Phys.}\ }%
  \textbf{\bibinfo {volume} {56}},\ \bibinfo {pages} {265--269} (\bibinfo
  {year} {1988})%
  \bibAnnoteFile{NoStop}{gruber88}%
\bibitem{sommerfeld52}%
  \BibitemOpen
  \bibfield{author}{%
  \bibinfo {author} {\bibfnamefont{A.}~\bibnamefont{Sommerfeld}},\ }%
  \emph{\bibinfo {title} {{Electrodynamics. Lectures on Theoretical
  Physics}}},\ Vol.\ \bibinfo {volume} {III}\ (\bibinfo {publisher} {Academic
  Press},\ \bibinfo {address} {New York},\ \bibinfo {year} {1952})%
  \bibAnnoteFile{NoStop}{sommerfeld52}%
\bibitem{schiff60}%
  \BibitemOpen
  \bibfield{author}{%
  \bibinfo {author} {\bibfnamefont{L.~I.}\ \bibnamefont{Schiff}},\ }%
  \bibfield{title}{%
  \enquote{\bibinfo {title} {{On Experimental Tests of the General Theory of
  Relativity}},}\ }%
  \bibfield{journal}{%
  \bibinfo {journal} {Am. J. Phys.}\ }%
  \textbf{\bibinfo {volume} {28}},\ \bibinfo {pages} {340--343} (\bibinfo
  {year} {1960})%
  \bibAnnoteFile{NoStop}{schiff60}%
\bibitem{rowlands97}%
  \BibitemOpen
  \bibfield{author}{%
  \bibinfo {author} {\bibfnamefont{P.}~\bibnamefont{Rowlands}},\ }%
  \bibfield{title}{%
  \enquote{\bibinfo {title} {{A simple approach to the experimental
  consequences of general relativity}},}\ }%
  \bibfield{journal}{%
  \bibinfo {journal} {Physics Education}\ }%
  \textbf{\bibinfo {volume} {32}},\ \bibinfo {pages} {49--55} (\bibinfo {year}
  {1997})%
  \bibAnnoteFile{NoStop}{rowlands97}%
\bibitem{cuzinatto11}%
  \BibitemOpen
  \bibfield{author}{%
  \bibinfo {author} {\bibfnamefont{R.~R.}\ \bibnamefont{Cuzinatto}}, \bibinfo
  {author} {\bibfnamefont{B.~M.}\ \bibnamefont{Pimental}},\ and\ \bibinfo
  {author} {\bibfnamefont{P.~J.}\ \bibnamefont{Pompeia}},\ }%
  \bibfield{title}{%
  \enquote{\bibinfo {title} {{Schwarzschild and de Sitter solutions from the
  argument by Lenz and Sommerfeld}},}\ }%
  \bibfield{journal}{%
  \bibinfo {journal} {Am. J. Phys.}\ }%
  \textbf{\bibinfo {volume} {79}},\ \bibinfo {pages} {662--667} (\bibinfo
  {year} {2011})%
  \bibAnnoteFile{NoStop}{cuzinatto11}%
\bibitem{rindler68}%
  \BibitemOpen
  \bibfield{author}{%
  \bibinfo {author} {\bibfnamefont{W.}~\bibnamefont{Rindler}},\ }%
  \bibfield{title}{%
  \enquote{\bibinfo {title} {{Counterexample to the Lenz-Schiff Argument}},}\
  }%
  \bibfield{journal}{%
  \bibinfo {journal} {Am. J. Phys.}\ }%
  \textbf{\bibinfo {volume} {36}},\ \bibinfo {pages} {540--544} (\bibinfo
  {year} {1968})%
  \bibAnnoteFile{NoStop}{rindler68}%
\bibitem{tangherlini62}%
  \BibitemOpen
  \bibfield{author}{%
  \bibinfo {author} {\bibfnamefont{F.~R.}\ \bibnamefont{Tangherlini}},\ }%
  \bibfield{title}{%
  \enquote{\bibinfo {title} {{Postulational Approach to Schwarz\-schild's
  Exterior Solution with Application to a Class of Interior Solutions}},}\ }%
  \bibfield{journal}{%
  \bibinfo {journal} {Nuovo Cimento}\ }%
  \textbf{\bibinfo {volume} {25}},\ \bibinfo {pages} {1081--1105} (\bibinfo
  {year} {1962})%
  \bibAnnoteFile{NoStop}{tangherlini62}%
\bibitem{sacks68}%
  \BibitemOpen
  \bibfield{author}{%
  \bibinfo {author} {\bibfnamefont{W.~M.}\ \bibnamefont{Sacks}}\ and\ \bibinfo
  {author} {\bibfnamefont{J.~A.}\ \bibnamefont{Ball}},\ }%
  \bibfield{title}{%
  \enquote{\bibinfo {title} {{Simple derivations of the Schwarzschild
  metric}},}\ }%
  \bibfield{journal}{%
  \bibinfo {journal} {Am. J. Phys.}\ }%
  \textbf{\bibinfo {volume} {36}},\ \bibinfo {pages} {240--245} (\bibinfo
  {year} {1968})%
  \bibAnnoteFile{NoStop}{sacks68}%
\bibitem{rindler69}%
  \BibitemOpen
  \bibfield{author}{%
  \bibinfo {author} {\bibfnamefont{W.}~\bibnamefont{Rindler}},\ }%
  \bibfield{title}{%
  \enquote{\bibinfo {title} {{Counterexample to the Tangherlini Argument}},}\
  }%
  \bibfield{journal}{%
  \bibinfo {journal} {Am. J. Phys.}\ }%
  \textbf{\bibinfo {volume} {37}},\ \bibinfo {pages} {72--73} (\bibinfo {year}
  {1969})%
  \bibAnnoteFile{NoStop}{rindler69}%
\bibitem{dadhich15}%
  \BibitemOpen
  \bibfield{author}{%
  \bibinfo {author} {\bibfnamefont{N.}~\bibnamefont{Dadhich}},\ }%
  \bibfield{title}{%
  \enquote{\bibinfo {title} {{Einstein is Newton with space curved}},}\ }%
  \bibfield{journal}{%
  \bibinfo {journal} {Current Science}\ }%
  \textbf{\bibinfo {volume} {109}},\ \bibinfo {pages} {260--264} (\bibinfo
  {year} {2015})%
  \bibAnnoteFile{NoStop}{dadhich15}%
\bibitem{mueller10}%
  \BibitemOpen
  \bibfield{author}{%
  \bibinfo {author} {\bibfnamefont{T.}~\bibnamefont{Mueller}}\ and\ \bibinfo
  {author} {\bibfnamefont{F.}~\bibnamefont{Grave}},\ }%
  \enquote{\bibinfo {title} {{Catalogue of Spacetimes}},}\ \bibinfo
  {howpublished} {arXiv:0904.4184v3 [gr-qc]} (\bibinfo {year} {2010})%
  \bibAnnoteFile{NoStop}{mueller10}%
\bibitem{jacobson07}%
  \BibitemOpen
  \bibfield{author}{%
  \bibinfo {author} {\bibfnamefont{T.}~\bibnamefont{Jacobson}},\ }%
  \bibfield{title}{%
  \enquote{\bibinfo {title} {{When is $g_{tt} g_{rr}=-1$?}}.}\ }%
  \bibfield{journal}{%
  \bibinfo {journal} {Class. Quant. Grav.}\ }%
  \textbf{\bibinfo {volume} {24}},\ \bibinfo {pages} {5717--5719} (\bibinfo
  {year} {2007})%
  \bibAnnoteFile{NoStop}{jacobson07}%
\bibitem{shuler18}%
  \BibitemOpen
  \bibfield{author}{%
  \bibinfo {author} {\bibfnamefont{R.}~\bibnamefont{Shuler}},\ }%
  \bibfield{title}{%
  \enquote{\bibinfo {title} {{A family of metric gravities}},}\ }%
  \bibfield{journal}{%
  \bibinfo {journal} {Eur. Phys. J. Plus}\ }%
  \textbf{\bibinfo {volume} {133}},\ \bibinfo {pages} {158} (\bibinfo {year}
  {2018})%
  \bibAnnoteFile{NoStop}{shuler18}%
\bibitem{gron07}%
  \BibitemOpen
  \bibfield{author}{%
  \bibinfo {author} {\bibfnamefont{\O.}\ \bibnamefont{Gr{\o}n}}\ and\ \bibinfo
  {author} {\bibfnamefont{S.}~\bibnamefont{Hervik}},\ }%
  \emph{\bibinfo {title} {{Einstein's General Theory of Relativity: With Modern
  Applications in Cosmology}}}\ (\bibinfo {publisher} {Springer Science \&
  Business Media},\ \bibinfo {address} {Springer, Berlin},\ \bibinfo {year}
  {2007})%
  \bibAnnoteFile{NoStop}{gron07}%
\bibitem{bunn09}%
  \BibitemOpen
  \bibfield{author}{%
  \bibinfo {author} {\bibfnamefont{E.~F.}\ \bibnamefont{Bunn}}\ and\ \bibinfo
  {author} {\bibfnamefont{D.~W.}\ \bibnamefont{Hogg}},\ }%
  \bibfield{title}{%
  \enquote{\bibinfo {title} {{The kinematic origin of the cosmological
  redshift}},}\ }%
  \bibfield{journal}{%
  \bibinfo {journal} {Am. J. Phys.}\ }%
  \textbf{\bibinfo {volume} {77}},\ \bibinfo {pages} {688--694} (\bibinfo
  {year} {2009})%
  \bibAnnoteFile{NoStop}{bunn09}%
\bibitem{davis04}%
  \BibitemOpen
  \bibfield{author}{%
  \bibinfo {author} {\bibfnamefont{T.~M.}\ \bibnamefont{Davis}}\ and\ \bibinfo
  {author} {\bibfnamefont{C.~H.}\ \bibnamefont{Lineweaver}},\ }%
  \bibfield{title}{%
  \enquote{\bibinfo {title} {{Expanding Confusion: common misconceptions of
  cosmological horizons and the superluminal expansion of the universe}},}\ }%
  \bibfield{journal}{%
  \bibinfo {journal} {Publ. Astron. Soc. Australia}\ }%
  \textbf{\bibinfo {volume} {21}},\ \bibinfo {pages} {97--109} (\bibinfo {year}
  {2004})%
  \bibAnnoteFile{NoStop}{davis04}%
\bibitem{schwarzschild16a}%
  \BibitemOpen
  \bibfield{author}{%
  \bibinfo {author} {\bibfnamefont{K.}~\bibnamefont{Schwarzschild}},\ }%
  \enquote{\bibinfo {title} {{\"Uber das Gravitationsfeld eines Massenpunktes
  nach der Einsteinschen Theorie}},}\ in\ \emph{\bibinfo {booktitle}
  {{Sitzungsberichte der K\"oniglich-Preu{\ss}ischen Akademie der
  Wissenschaften}}}\ (\bibinfo {publisher} {Reimer},\ \bibinfo {address}
  {Berlin},\ \bibinfo {year} {1916})\ pp.\ \bibinfo {pages} {189--196},\
  \bibinfo {note} {{English translation: \emph{On the Gravitational Field of a
  Mass Point According to Einstein's Theory}, S. Antoci and A. Loinger,
  arXiv:physics/9905030v1}}%
  \bibAnnoteFile{NoStop}{schwarzschild16a}%
\bibitem{painleve21}%
  \BibitemOpen
  \bibfield{author}{%
  \bibinfo {author} {\bibfnamefont{P.}~\bibnamefont{Painlevé}},\ }%
  \bibfield{title}{%
  \enquote{\bibinfo {title} {{La mécanique classique et la théorie de la
  relativité}},}\ }%
  \bibfield{journal}{%
  \bibinfo {journal} {{C. R. Acad. Sci. (Paris)}}\ }%
  \textbf{\bibinfo {volume} {173}},\ \bibinfo {pages} {677--680} (\bibinfo
  {year} {1921})%
  \bibAnnoteFile{NoStop}{painleve21}%
\bibitem{gullstrand22}%
  \BibitemOpen
  \bibfield{author}{%
  \bibinfo {author} {\bibfnamefont{A.}~\bibnamefont{Gullstrand}},\ }%
  \bibfield{title}{%
  \enquote{\bibinfo {title} {{Allgemeine Lösung des statischen
  Einkörperproblems in der Einsteinschen Gravitationstheorie}},}\ }%
  \bibfield{journal}{%
  \bibinfo {journal} {{Arkiv. Mat. Astron. Fys.}}\ }%
  \textbf{\bibinfo {volume} {16}},\ \bibinfo {pages} {1--15} (\bibinfo {year}
  {1922})%
  \bibAnnoteFile{NoStop}{gullstrand22}%
\bibitem{harvey06}%
  \BibitemOpen
  \bibfield{author}{%
  \bibinfo {author} {\bibfnamefont{A.}~\bibnamefont{Harvey}}, \bibinfo {author}
  {\bibfnamefont{E.}~\bibnamefont{Schucking}},\ and\ \bibinfo {author}
  {\bibfnamefont{E.~J.}\ \bibnamefont{Surowitz}},\ }%
  \bibfield{title}{%
  \enquote{\bibinfo {title} {{Redshifts and Killing vectors}},}\ }%
  \bibfield{journal}{%
  \bibinfo {journal} {Am. J. Phys.}\ }%
  \textbf{\bibinfo {volume} {74}},\ \bibinfo {pages} {1017--1024} (\bibinfo
  {year} {2006})%
  \bibAnnoteFile{NoStop}{harvey06}%
\bibitem{rindler01}%
  \BibitemOpen
  \bibfield{author}{%
  \bibinfo {author} {\bibfnamefont{W.}~\bibnamefont{Rindler}},\ }%
  \emph{\bibinfo {title} {{Relativity. Special, general, and cosmological}}}\
  (\bibinfo {publisher} {Oxford Univ. Press},\ \bibinfo {address} {New York},\
  \bibinfo {year} {2001})%
  \bibAnnoteFile{NoStop}{rindler01}%
\bibitem{okun99}%
  \BibitemOpen
  \bibfield{author}{%
  \bibinfo {author} {\bibfnamefont{L.~B.}\ \bibnamefont{Okun'}}, \bibinfo
  {author} {\bibfnamefont{K.~G.}\ \bibnamefont{Selvanov}},\ and\ \bibinfo
  {author} {\bibfnamefont{V.L.}\ \bibnamefont{Telegdi}},\ }%
  \bibfield{title}{%
  \enquote{\bibinfo {title} {{Gravitation, photons, clocks}},}\ }%
  \bibfield{journal}{%
  \bibinfo {journal} {{Physics -- Uspekhi}}\ }%
  \textbf{\bibinfo {volume} {42}},\ \bibinfo {pages} {1045 -- 1050} (\bibinfo
  {year} {1999})%
  \bibAnnoteFile{NoStop}{okun99}%
\bibitem{baez05}%
  \BibitemOpen
  \bibfield{author}{%
  \bibinfo {author} {\bibfnamefont{J.~C.}\ \bibnamefont{Baez}}\ and\ \bibinfo
  {author} {\bibfnamefont{E.~F.}\ \bibnamefont{Bunn}},\ }%
  \bibfield{title}{%
  \enquote{\bibinfo {title} {{The meaning of Einstein's equation}},}\ }%
  \bibfield{journal}{%
  \bibinfo {journal} {Am. J. Phys.}\ }%
  \textbf{\bibinfo {volume} {73}},\ \bibinfo {pages} {644--652} (\bibinfo
  {year} {2005})%
  \bibAnnoteFile{NoStop}{baez05}%
\bibitem{brans61}%
  \BibitemOpen
  \bibfield{author}{%
  \bibinfo {author} {\bibfnamefont{C.}~\bibnamefont{Brans}}\ and\ \bibinfo
  {author} {\bibfnamefont{R.~H~.}\ \bibnamefont{Dicke}},\ }%
  \bibfield{title}{%
  \enquote{\bibinfo {title} {{Mach's Principle and a Relativistic Theory of
  Gravitation}},}\ }%
  \bibfield{journal}{%
  \bibinfo {journal} {Phys. Rev.}\ }%
  \textbf{\bibinfo {volume} {124}},\ \bibinfo {pages} {925--935} (\bibinfo
  {year} {1961})%
  \bibAnnoteFile{NoStop}{brans61}%
\bibitem{weisberg10}%
  \BibitemOpen
  \bibfield{author}{%
  \bibinfo {author} {\bibfnamefont{J.~M.}\ \bibnamefont{Weisberg}}, \bibinfo
  {author} {\bibfnamefont{D.~J.}\ \bibnamefont{Nice}},\ and\ \bibinfo {author}
  {\bibfnamefont{J.~H.}\ \bibnamefont{Taylor}},\ }%
  \bibfield{title}{%
  \enquote{\bibinfo {title} {{Timing Measurements of the Relativistic Binary
  Pulsar PSR B1913+16}},}\ }%
  \bibfield{journal}{%
  \bibinfo {journal} {Astrophys. J.}\ }%
  \textbf{\bibinfo {volume} {722}},\ \bibinfo {pages} {1030--1034} (\bibinfo
  {year} {2010})%
  \bibAnnoteFile{NoStop}{weisberg10}%
\bibitem{abbott16}%
  \BibitemOpen
  \bibfield{author}{%
  \bibinfo {author} {\bibfnamefont{B.~P.~Abbott}\ \bibnamefont{et~al.}},\ }%
  \bibfield{title}{%
  \enquote{\bibinfo {title} {{Observation of Gravitational Waves from a Binary
  Black Hole Merger}},}\ }%
  \bibfield{journal}{%
  \bibinfo {journal} {Phys. Rev. Lett.}\ }%
  \textbf{\bibinfo {volume} {116}},\ \bibinfo {pages} {061102} (\bibinfo {year}
  {2016})%
  \bibAnnoteFile{NoStop}{abbott16}%
\bibitem{abbott18}%
  \BibitemOpen
  \bibfield{author}{%
  \bibinfo {author} {\bibfnamefont{B.~P.~Abbott}\ \bibnamefont{et~al.}},\ }%
  \enquote{\bibinfo {title} {{GWTC-1: A Gravitational-Wave Transient Catalog of
  Compact Binary Mergers Observed by LIGO and Virgo during the First and Second
  Observing Runs}},}\ \bibinfo {howpublished} {arXiv:1811.12907v2
  [astro-ph.HE]} (\bibinfo {year} {2018})%
  \bibAnnoteFile{NoStop}{abbott18}%
\end{thebibliography}
\end{document}